\documentclass[fleqn,usenatbib]{mnras}

\usepackage{fix-cm}      
\usepackage{newtxtext,newtxmath}
\usepackage{bm}           

\usepackage{amsmath} 
\usepackage{multirow}
\usepackage{adjustbox}
\usepackage{float}
\usepackage{caption}
\usepackage{subcaption}
\usepackage{array}
\usepackage{makecell}   
\usepackage[T1]{fontenc}
\usepackage{graphicx}		
\usepackage{booktabs}
\usepackage{hyperref}
\usepackage[nameinlink]{cleveref}
\usepackage{orcidlink}
\usepackage{CJKutf8}
\usepackage[normalem]{ulem}

\newcommand{\kepler}{{\em Kepler\/}}
\newcommand{\corot}{{\em CoRoT\/}}
\newcommand{\tess}{{\em TESS\/}}
\newcommand{\gaia}{{\em Gaia\/}}
\newcommand{\echelle}{\'{e}chelle}

\newcommand{\ac}[1]{{{\textcolor{black}{#1}}}}

\crefname{equation}{equation}{equations}

\newcommand{\eqnref}[1]{%
  \textup{equation~(\ref{#1})}%
}
\newcommand{\figref}[1]{%
  \textup{Fig.~\ref{#1}}%
}
\newcommand{\tabref}[1]{%
  \textup{Table~\ref{#1}}%
}
\newcommand{\secref}[1]{%
  \textup{section~\ref{#1}}%
}

\title[Secondary Clump Star $\kappa$~Cyg]{CHARA Interferometry and \tess{} Asteroseismology of the Core-Helium Burning Red Giant $\mathbf{\kappa}$~Cyg}

\newcommand{\CNnames}[1]{{\begin{CJK}{UTF8}{gbsn}(#1)\end{CJK}}}

\author[Chowhan et al.]{Arnab Chowhan\orcidlink{0009-0006-5394-8331}$^{1}$\thanks{E-mail: acho0151@uni.sydney.edu.au},
Timothy R. Bedding\orcidlink{0000-0001-5222-4661}$^{1}$,
Daniel Huber\orcidlink{0000-0001-8832-4488}$^{2}$,
J. M. Joel Ong \CNnames{王加冕}\orcidlink{0000-0001-7664-648X}$^{1}$,
\newauthor Lea S. Schimak\orcidlink{0009-0006-8575-2106}$^{1}$,
Yaguang Li \CNnames{李亚光}\orcidlink{0000-0003-3020-4437}$^{2}$,
Courtney L. Crawford\orcidlink{0000-0002-7654-7438}$^{1}$,
Timothy R. White\orcidlink{0000-0002-6980-3392}$^{3}$\\
$^{1}$Sydney Institute for Astronomy, School of Physics, University of Sydney,
Sydney, NSW 2006, Australia\\
$^{2}$Institute for Astronomy, University of Hawai‘i,
2680 Woodlawn Drive, Honolulu, HI 96822, USA\\
$^{3}$Sydney Informatics Hub, Core Research Facilities, University of Sydney, NSW 2006, Australia 
}

\date{Accepted 2026 April 13. Received 2026 April 10; in original form 2026 March 12}

\pubyear{\the\year{}}

\begin{document}
\label{firstpage}
\pagerange{\pageref{firstpage}--\pageref{lastpage}}
\maketitle

\begin{abstract}
We present a detailed study of the secondary red clump star, $\kappa$~Cyg, by combining long-baseline visible interferometry using the PAVO beam combiner at the CHARA Array with high-precision asteroseismology from \tess{}. This dual approach allowed for a stringent test of stellar evolutionary models in the core helium-burning phase, which remains a regime of significant theoretical uncertainty. Using the PAVO interferometric data and fitting the limb-darkened intensity profile directly, we measured $R = 8.65\pm0.10~\rm R_\odot$. We fitted the spectral energy distribution (SED) using Phoenix model atmospheres and calculated $L = 44.46 \pm 1.09~\rm L_\odot$ and $T_{\rm eff} = 5066^{+47}_{-50}~\mathrm{K}$. Using 16 sectors of \tess{} photometry, we detected clear solar-like oscillations in $\kappa$~Cyg. Through comparison of oscillation frequencies with MESA grids using either predictive mixing (PM) or exponential overshooting (OS), we found that models reproducing the oscillation frequencies systematically overestimate the stellar radius, with overshooting models performing only marginally better. The same models also under-predict the observed dipole-mode period spacing ($\Delta\Pi_1$). 
By inspecting the phase offset ($\epsilon_p$), we conclude that models misrepresent the interior structure of the star. 
Our results demonstrate that matching envelope-dominated asteroseismic observables alone is insufficient to ensure a correct core or even global structure, and highlight the need for improved treatments of convective boundary mixing in the models of core helium-burning (CHeB) stars.
\end{abstract}

\begin{keywords}
asteroseismology -- stars: fundamental parameters -- stars: horizontal branch -- stars: oscillations -- techniques: interferometric 
\end{keywords}



\section{Introduction}
\label{intro}

Asteroseismology and long-baseline interferometry are, separately, already powerful observational techniques. Combined, they have the potential ability to determine fundamental stellar properties with a level of precision that neither technique can achieve in isolation. This synergy provides a powerful framework for testing stellar models by linking probes of stellar interiors with independent constraints on stellar size and effective temperature.

Over the past two decades, space-based photometry from \ac{\corot{} \citep{Baglin2006a, Baglin2006b}, \kepler{} \citep{Borucki2010} and \tess{} \citep{Ricker2015}}, complemented by ground-based radial velocities \citep[e.g.,][]{Campante2024, lundkvist2024, Li2025}, has \ac{enabled high-precision study of stellar oscillations. These data have established asteroseismology as} a mature observational discipline, delivering precise masses, radii, and evolutionary states for thousands of stars \citep[e.g.,][]{chaplin2013, hekker2017, Garcia2019, Jackiewicz2021, Aerts2021, Huber2025}. Yet precision alone is not sufficient: the radii and masses inferred from asteroseismic modelling are only as reliable as the physical assumptions built into the stellar evolution models themselves. Asteroseismic radius, rather than being a direct measurement of a star's physical size, corresponds to the radius of the stellar model whose computed mode frequencies best reproduce the observed oscillation spectrum. Therefore, it is sensitive to the adopted prescriptions for interior physics, chemical mixing, and the treatment of near-surface layers. 

This model-dependence is most acute for stars undergoing core helium burning (CHeB). Once a star exhausts hydrogen in its core and ignites helium, it develops strong chemical gradients at multiple locations in its interior, governed by convective boundary mixing, one of the most uncertain processes in stellar physics. Secondary red clump stars, which are intermediate-mass stars burning helium in a non-degenerate core, are valuable targets for testing these prescriptions. They occupy a narrow region of the Hertzsprung–Russell diagram, overlapping with both primary red clump stars and ascending red giant branch stars \citep{Girardi1999, girardi2016}, making them difficult to identify photometrically. Despite their potential, CHeB stars have received comparatively little focused asteroseismic attention \citep[e.g.,][]{Moura2020, murphy2021, brogaard2023}, with most modelling efforts concentrated on reproducing their global properties rather than their individual mode frequencies \citep[e.g.,][]{Bossini2015, bossini2017, constantino2015, Noll2023, Noll2025}.

An independent and complementary measurement of stellar properties is provided by long-baseline interferometry. Unlike asteroseismology, interferometry determines angular diameters directly from the spatial coherence of the stellar disk, yielding radii that are largely independent of assumptions about interior physics \citep[e.g.,][]{Quirrenbach1996}. However, a weak model-dependence does enter through the limb-darkening coefficients required to convert the observed stellar visibility into an angular diameter. By sampling interferometric measurements beyond the first null \ac{in the stellar visibility curve}, where the curve is most sensitive to the intensity profile of the stellar disk, one can reduce reliance on hydrodynamic atmosphere models and even use the data to test the models directly \citep[e.g.,][]{Kervella2017}. 

Combined with precise parallaxes from \gaia{} \citep{Gaia2022} and bolometric fluxes \citep[e.g.,][]{Boyajian2014}, interferometry delivers direct measurements of stellar radii and effective temperatures. The complementarity of the two techniques has been well established across the H--R diagram \citep[e.g.,][]{North2007, Mazumdar2009, Huber2012, White2013, johnson2014, horringgaard2017, Stokholm2019, Li2025}. This makes interferometry not merely a tool for improving parameter precision, but a genuine test of stellar structure theory, one that we exploit here for the first time in a secondary red clump star.

In this work, we present the first combined analysis of interferometric observations and detailed asteroseismic modelling of individual oscillation modes for a secondary red clump star. By using interferometric measurements beyond the first null \ac{in the visibility curve}, we directly test hydrodynamic atmosphere models and the limb-darkening coefficients they predict. We then compare the resulting stellar radii and effective temperatures against asteroseismic model predictions to investigate how well current stellar models, with their prescribed treatments of convective boundary mixing and near-surface layers, reproduce the internal and external structure of a core helium burning star, with implications for our understanding of mixing in evolved stars.

\section{\texorpdfstring{Properties of $\kappa$~Cyg}{Properties of kappa Cyg}}
\label{properties}
We target the bright G8 red giant $\kappa$~Cyg (HR~7328; HD~181276; HIP~94779) for several practical and scientific reasons. It is a nearby evolved star (parallax $\varpi_{Gaia}=26.49$~mas) whose brightness ($V=3.76$) makes it an excellent target for long-baseline interferometric observations. At the same time, it is not bright enough to saturate the \tess{} detectors and produce long bleeding columns in the target pixel files (TPFs), which would require special techniques to extract high-quality photometry \citep{Rudrasingam2026}. The resulting asteroseismic data allow a clear identification of its evolutionary state using established diagnostics based on dipolar mixed modes (e.g. \citealt{bedding2011, mosser2012}). As a secondary clump star, $\kappa$~Cyg therefore provides a rare opportunity to obtain both interferometric and asteroseismic constraints on a star in this evolutionary phase.

Numerous spectroscopic studies have analysed $\kappa$~Cyg and provide well-constrained atmospheric parameters. We adopted the effective temperature, $T_{\rm eff}=5021\pm100$~K, surface gravity, $\log{g}=3.02\pm0.08$ and metallicity, $\rm [Fe/H]=0.10\pm0.07$ from \cite{deka2018}, with uncertainties chosen to be conservative and consistent with typical spectroscopic analyses \citep{bruntt2010, Tayar2022}. These parameters are used in \secref{vis-model} to compute the limb-darkened angular diameter ($\theta_{\rm LD}$) using 3D STAGGER atmosphere models.

$\kappa$~Cyg has previously been studied using NPOI (Navy Prototype Optical Interferometer; \citealt{Benson2003}) and its interferometric $\theta_{\rm LD}$ was measured to be $2.143 \pm 0.008$~mas \citep{Baines2018}. In the present work, we employ new visible-wavelength interferometric observations from the PAVO beam combiner at the CHARA Array, with visibility measurements extending well beyond the first null, allowing a more robust and less model-dependent determination of the stellar intensity profile.

The \gaia{} DR3 solution for $\kappa$~Cyg yields a RUWE (re-normalised unit weight error) of 2.29, which indicates excess astrometric noise that is usually ascribed to binarity \citep{Ginard2024}. However, no evidence for binarity has been reported in the literature, and no signatures of a companion are detected in the asteroseismic power spectrum (see \secref{global-params}). \ac{It has been shown that, for a given separation, a low RUWE corresponds to a larger magnitude difference between the primary and secondary components in the binary (see fig. 8 of \citealt{Belokurov2020}). So, even if $\kappa$~Cyg were to be detected as a binary, its lower RUWE indicates that the probable companion has low luminosity. Since the photon count is negligible from the probable companion, the visibilities and hence interferometric results should not be impacted.} We therefore treat $\kappa$~Cyg as a single star in the analysis that follows.

We did not employ any Doppler correction to our power spectrum since
the radial velocity of $\kappa$~Cyg is $29.36 \pm 0.03~\mathrm{km\ s^{-1}}$
\citep{deka2018}, corresponding to a frequency shift of $\sim 0.01~\mu$Hz, which is well below the observational uncertainties in the frequency range of interest.

\section{Interferometry}

\subsection{Observations}
\label{inter-obs}
We obtained interferometric observations using the Precision Astronomical Visible Observations (PAVO) beam combiner at the Center for High Angular Resolution Astronomy (CHARA) Array \citep{Brummelaar2005} at Mount Wilson Observatory, California. PAVO is a pupil-plane beam combiner operating mainly in the visible spectrum ($630$--$900$~nm), spectrally dispersing light across several independent channels. The CHARA Array provides 15 possible baselines, with the longest being the S1E1 telescope configuration (330~m). Operating at visible wavelengths grants PAVO high spatial resolution and it has a limiting magnitude of approximately $R \sim 7$--8, depending on seeing conditions. A detailed description of the instrument can be found in \cite{Ireland2008}.

\begin{figure}
\includegraphics[width=\columnwidth]{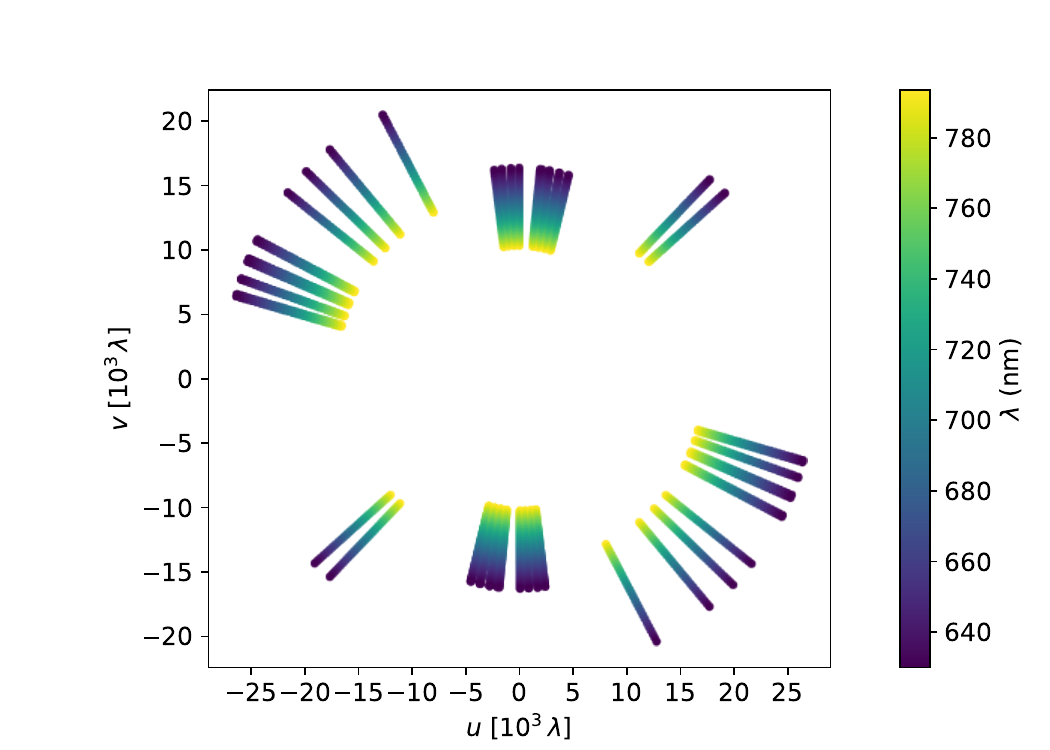}
  \caption{Spatial frequency coverage of $\kappa$~Cyg with 7 nights of PAVO observations in two-telescope mode. The points are colour-coded by wavelength, with bluer corresponding to lower wavelengths, shown in the colour map on the right.}
     \label{fig:uvPlot}
\end{figure}

We used PAVO in two-telescope mode over seven nights between 2018~April and 2020~July.  Observations were performed using the W1W2 (107.93~m) and E1E2 (65.88~m) baselines, selected to sample both within and beyond the first null of the stellar visibility function. The complete log of our observations, including dates, telescope configurations and calibrators, is provided in \tabref{tab:interferometric_observations}. The corresponding coverage in the spatial frequency plane is illustrated in \figref{fig:uvPlot}. The points are colour-coded by wavelength channel, demonstrating the broad $(u, v)$ coverage from only two baselines.

\begin{table*}
    \centering
    \renewcommand{\arraystretch}{1.2} 
    \setlength{\tabcolsep}{8pt} 
    \begin{tabular}{c c c c c}
        \hline
        UT Date & No. of Observations & Telescopes & Baseline & Calibrator Stars \\
        \hline
         2018 April 22 & 4 & W1W2 & 107.93 & HD 177003, HD 188252\\ 
        2018 April 23 & 4 & W1W2 & 107.93 & HD 177003, HD 188252\\
        2018 June 02 & 4 & W1W2 & 107.93 & HD 177003, HD 188252\\
        2018 August 05 & 4 & E1E2 & 65.88& HD 177003, HD 188252, HD 175640\\
        2019 August 24 & 4 & E1E2 & 65.88& HD 177003\\
        2019 August 25 & 2 & W1W2 & 107.93 & HD 177003\\
        2020 July 22 & 4 & E1E2 & 65.88 & HD 177003, HD 188252\\
        \hline
    \end{tabular}
    \caption{Log of interferometric observations for $\kappa$~Cyg.}
    \label{tab:interferometric_observations}
\end{table*}

For each scan, we observed calibrators immediately before and after observing our target to track the interferometric transfer function. A bright, unresolved point source close to the target object is ideal for calibration to ensure spatially and temporally similar observations. We selected \ac{bright} B stars as calibrators \ac{because they} have angular diameters ($\leq0.2$~mas) too small to be resolved by PAVO. The full list of calibrators is provided in \tabref{tab:calibrator_stars}. We also observed $\iota^1$~Cyg (HD~183534) but excluded it from the list of calibrators since its raw visibilities showed structure indicative of a probable companion.

The expected angular sizes of the calibrators ($\theta_{V-K}$) were calculated using the empirical $V - K$ surface-brightness relation from \citet{Boyajian2014}. We took $V$ magnitudes from the Tycho catalogue \citep{Ammons2006} and converted to the Johnson system following \citet{Bessell2000}, while $K$ magnitudes were taken from the Two Micron All Sky Survey (2MASS; \citealt{Skrutskie2006}). We estimated the interstellar reddening for each calibrator using the 3D dust maps from \citet{Green2019} and followed the extinction law from \citet{Donnell1994} to deredden the magnitudes. 

We have used the 28 central wavelength channels for our analysis and employed the PAVO data-reduction pipeline, which has been used for numerous two-telescope visibility studies \citep[e.g.,][]{Bazot2011, Derekas2011, Huber2012, White2013, Maestro2013}. The calibrated squared-visibility measurements for $\kappa$~Cyg as a function of spatial frequency are presented in \figref{fig:visibility}. The data points, from a total of 26 scans, are colour-coded by wavelength channel, consistent with \figref{fig:uvPlot}.

\begin{table}
    \centering
    \renewcommand{\arraystretch}{1.2} 
    \setlength{\tabcolsep}{8pt} 
    \begin{tabular}{l c c c c}
        \hline
        HD & Sp. T. & $V - K$ & $E_{B - V}$ & $\theta_{V - K}$ (mas) \\
        \hline
        177003 & B2.5IV & -0.518 & 0.000 & 0.209\\
        188252 & B2III & -2.807 & 0.090 & 0.047\\
        175640 & B9III & -4.095 & 0.871 & 0.055\\
        \hline
    \end{tabular}
    \caption{Calibrators observed by PAVO for $\kappa$~Cyg and their properties.}
    \label{tab:calibrator_stars}
\end{table}

\subsection{Visibility Modelling}
\label{vis-model}

\begin{figure*}
    \centering
    \includegraphics[width=\textwidth]{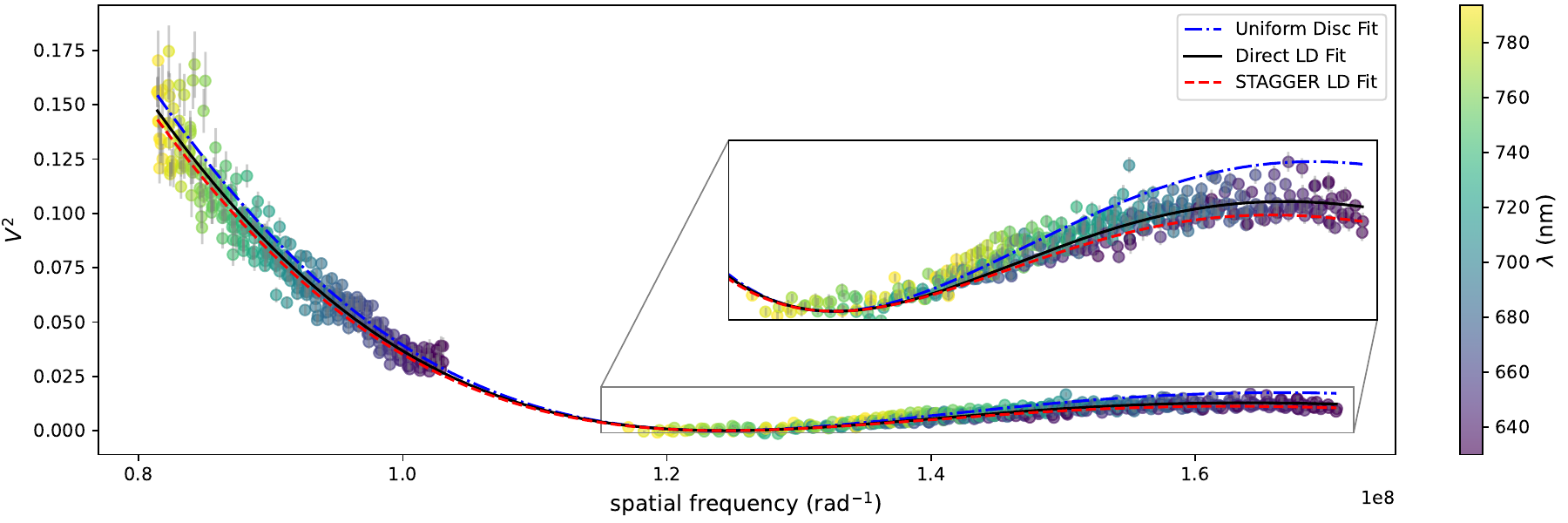}  
    \caption{Squared visibility of $\kappa$~Cyg as a function of spatial frequency, colour-coded by wavelength. The blue dash-dotted line shows the uniform-disc (UD) model \ac{($\chi^2_{\rm red}$=14.27)}, the black solid line shows the direct limb-darkening (LD) fit \ac{($\chi^2_{\rm red}$=5.26)}, and the red dashed line shows the STAGGER grid LD fit \ac{($\chi^2_{\rm red}$=6.83)}. The inset is shown for visual clarity.}
    \label{fig:visibility}
\end{figure*}

\begin{figure}
    \centering
    \includegraphics[width=0.47\textwidth]{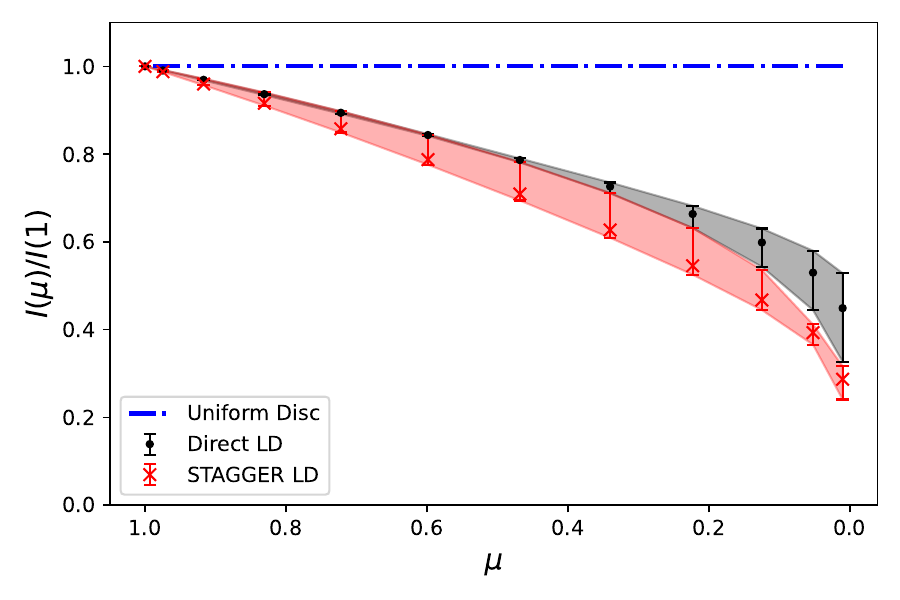}
    \caption{ Centre-to-limb intensity variation corresponding to the three cases shown in \figref{fig:visibility}.}
    \label{fig:clv}
\end{figure}
    
We illustrate the need to account for the centre-to-limb intensity variation in deriving an angular diameter ($\theta$) by first fitting a notional uniform-disc (UD) model. Although this simple formulation fits reasonably well up to the first null \ac{in the visibility curve}, it fails to reproduce the amplitude of the second lobe, which can be seen in the blue dash-dotted line of \figref{fig:visibility}. This discrepancy confirms the need to account for limb darkening (LD), the physical phenomenon in which the stellar disc intensity drops off towards the limb due to geometrical and atmospheric effects. The fitted uniform-disc angular diameter ($\theta_{\rm UD}$) nonetheless serves as a lower bound on the true angular diameter.

To model the visibility data more accurately, particularly in the higher spatial frequency regime where LD effects are pronounced, we employed the four-term non-linear LD intensity profile, $I(\mu)$, introduced by \citet{Claret2000}:
\begin{equation}
    \frac{I(\mu)}{I(1)} = 1 - \sum_{k=1}^4 a_k\left(1 - \mu^{k / 2}\right).
    \label{eq:limb_darkening_polynomial}
\end{equation}
Here, $\mu = \cos{\gamma}$, $\gamma$ is the angle between the line of sight and the stellar surface normal, and $a_k$ are the LD coefficients. This formulation improves upon the simple linear LD profile \citep{Schwarzschild1906}, 
\begin{equation}
    \dfrac{I(\mu)}{I(1)} = 1 -u(1-\mu),
    \label{eq:limb_darkening_linear}
\end{equation}
which could not sufficiently reproduce the models of \citet{Claret2011} and the intensity profiles observed by \citet{Klinglesmith1970}.

The van~Cittert--Zernike theorem \citep{Cittert1934, Zernike1938} relates the intensity distribution of a source to its corresponding fringe visibility via a Fourier transform. Following the approach of \citet{Quirrenbach1996} for a generalized polynomial LD law, the visibility corresponding to \eqnref{eq:limb_darkening_polynomial} is calculated as
\begin{equation}
    \begin{aligned}
    V(x) =2
    \dfrac{
    \begin{aligned}
    \left(1-\sum_{k=1}^4 a_k\right)\frac{J_1(x)}{x}
    + \sum_{k=1}^4 a_k 2^{k/4}\Gamma\!\left(\frac{k}{4}+1\right)
    \frac{J_{k/4+1}(x)}{x^{k/4+1}}
    \end{aligned}
    }{
    \begin{aligned}
  \left(
            1-\sum_{k=1}^4 \dfrac{ka_k}{k+4}\right)
    \end{aligned}
    }.
    \end{aligned}
    \label{eq:visibility_polynomial}
\end{equation}
Here, $x = \pi\theta_{\rm LD} \dfrac{B}{\lambda}$, with $B$ the projected baseline, $\theta_{\rm LD}$ the limb-darkened angular diameter, $\lambda$ the observing wavelength,  $\Gamma(z)$ the gamma function, and $J_n(x)$ the nth-order ordinary Bessel function of the first kind. The quantity $\dfrac{B}{\lambda}$ is the spatial frequency.

We tested two approaches to determining $\theta_{\rm LD}$. First, we used the STAGGER grid of \textit{ab initio} 3D hydrodynamic stellar atmosphere models \citep{Magic2013} to obtain theoretical LD coefficients, $a_k$ \citep{Magic2015}, based on the spectroscopic parameters of $\kappa$~Cyg (see \secref{properties}). Consistent with previous studies \citep[e.g.,][]{Huber2012, Karovicova2020}, we adopted coefficients corresponding to the Johnson $R$ band ($\approx 670$~nm), which is close to the central PAVO wavelength. Using the STAGGER coefficients, we fit our visibility data to \eqnref{eq:visibility_polynomial} to determine $\theta_{\rm LD}$. This is shown in \figref{fig:visibility} as a red dashed line. Although it fits significantly better than the UD prescription, it underestimates the amplitude of the second lobe, indicating that the models predict stronger limb darkening than is observed.

Second, as an alternative to using the 3D models, we performed a direct fit of \eqnref{eq:visibility_polynomial} to the observations, treating the $a_k$ and $\theta_{\rm LD}$ as five free parameters. To ensure the physical realism of the derived coefficients, we enforced three constraints that are based on the expected behaviour of stellar intensity:
\begin{itemize}
    \item the intensity across the entire stellar disc must be positive: $I(0\le\mu\le 1) > 0$.
    \item the intensity must decrease monotonically from the centre to the limb, i.e. $I(\mu)$ is an increasing function, $\frac{\partial I}{\partial\mu} > 0$.
    \item the rate of intensity drop should be sharper near the limb than at the centre, i.e.  $\frac{\partial^2 I}{\partial\mu^2} \leq 0$.
\end{itemize}
This constrained fit successfully reproduces the visibility amplitudes in the second lobe (black solid line in \figref{fig:visibility}). We compare the centre-to-limb intensity variations ($I(\mu)/I(1)$) corresponding to our direct fit and the STAGGER-based fit in \figref{fig:clv}, with the UD shown for reference. The direct fit exhibits weaker LD, resulting in a better match to the high spatial-frequency visibility data. 

We present the angular diameters and physical radii ($R$) from the three methods in \tabref{tab:interResults}. Following \citet{Cunha2007}, we calculated the radius using the \gaia{} DR3  parallax \citep{Gaia2022}, $\varpi$, as
\begin{equation}
    \dfrac{R}{\rm R_\odot}=107.47\dfrac{\theta_{\rm LD}}{\varpi}
    \label{eq:radius-parallax}
\end{equation}
For completeness, we also mention the results from using a linear LD profile (i.e. \eqnref{eq:limb_darkening_linear}). The stronger LD predicted by the STAGGER grid produces $R$ approximately 2.1 per cent higher than the direct fit. Because high spatial frequency visibilities are very sensitive to the intensity profile, directly adopting theoretical coefficients without sampling the second lobe can systematically bias the radius measurements \citep{Kervella2017}. Similar discrepancies in the centre-to-limb intensity profiles have been noted in transit studies of dwarfs, and have been attributed to not including the magnetic induction equation in the hydrodynamic models \citep{Maxted2023, Kostogryz2024, Verma2024}. We plan to explore this systematic difference of models predicting higher LD in a forthcoming study of other targets that have visibility measurements from PAVO beyond the first null.

\subsection{Bolometric Flux}
\label{fbol}

\begin{figure}
    \includegraphics[width=\columnwidth]{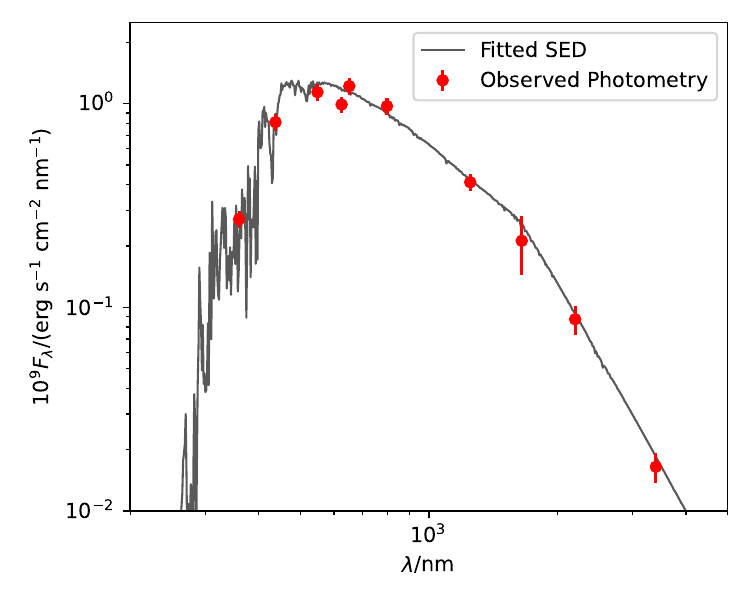}
    \caption{Spectral energy distribution of $\kappa$~Cyg based on available spectrophotometry, shown as red dots. The best-fit PHOENIX spectral template is shown in dark grey.}
    \label{fig:sed}
\end{figure}

 We determined the bolometric flux ($F_{\rm bol}$) and luminosity ($L$) of $\kappa$~Cyg by fitting its spectral energy distribution (SED). We compiled a comprehensive set of spectrophotometric measurements from \citealt{Ducati2002, Cutri2003, Gaia2022}. The adopted magnitudes and corresponding fluxes are listed in \tabref{tab:stellar_mag}. We assigned an uncertainty of $0.05$~mag to the optical bands and employed the reported uncertainties from the sources for the remaining bands. To compare observations with theory, we used PHOENIX stellar atmosphere models \citep{Allard2011, Allard2012}. We integrated these model fluxes over the response curves of each filter to produce synthetic fluxes. The filter transmission functions and zero points were taken from \citet{Bessell2000} for $UBVRI$, \citet{Bessell1998} for $JHKL$ and \citet{Evans2018} for \gaia{} $G$ band. To account for interstellar extinction, we used the wavelength-dependent reddening relations from \cite{Cardelli1989}, and the reddened synthetic fluxes were fitted to the observed ones. The best fit (with reddening, $A_v=0.03$ mag) is shown in dark grey in \figref{fig:sed}. \ac{We chose not to use PHOENIX models to derive LD coefficients for our analysis, since 3D radiative hydrodynamic models provide a more realistic description of centre-to-limb intensity variations than 1D atmospheres, owing to their explicit treatment of convective motions \citep[e.g.,][]{Asplund2009,Hayek2012, Magic2015}. Stronger LD is predicted from PHOENIX models \citep{Claret2014} compared to STAGGER, leading to a worse fit with $\theta_{\rm LD}=2.213^{+0.002}_{-0.005}$~mas and $\chi^2_{\rm red}\sim9.7$. Also, using two different models (PHOENIX and STAGGER) for $T_{\rm eff}$ and $\theta_{\rm LD}$ does not affect our results because the linear LD coefficient from STAGGER changes by $\sim0.02$ per 100~K, which is well within our uncertainties.}

Next, we integrated the fitted SED over wavelength using the trapezoidal rule to obtain $F_{\rm bol}$. Since $\kappa$~Cyg has very low flux outside the region covered by PHOENIX spectra ($200$--$3000$~nm), we \ac{followed the standard approach \citep[e.g.,][]{Mann2013,Ruffle2015} of extrapolating} the fitted spectrum following Wien's approximation in the bluer part and Rayleigh-Jeans law in the redder part. The uncertainty on $F_{\rm bol}$ includes systematic effects associated with photometric calibration, extrapolation of fitted SED beyond the observed wavelength range, photometric errors, and interstellar extinction. Therefore, we determined an uncertainty on $F_{\rm bol}$ using Monte Carlo resampling, where the relevant parameters were perturbed within their respective uncertainties, and found a value of 2.3~per cent. Our calculated flux is $F_{\rm bol} = (9.98 \pm 0.24) \times 10^{-7}\,\mathrm{erg\,s^{-1}\,cm^{-2}}$. We adopt this value for our analysis. We cross-checked our calculation with the stellar spectral atlas of \citet{Pickles1998} and obtained $F_{\rm bol} = (9.71 \pm 0.24) \times 10^{-7}\,\mathrm{erg\,s^{-1}\, cm^{-2}}$, which is lower by around 2.7~per cent. A similar offset in the same direction was observed for late K and M dwarfs by \citet{Mann2013}.

We calculated luminosity using the inverse square relation, $L= 4\pi  d^2 F_ {\rm bol}$, where $d$ is the distance of the star, calculated as the inverse of parallax, $\varpi$. This yields $L = (44.46\pm1.09)\ L_\odot$.

Once we have $\theta_ {\rm LD}$, we can find the effective temperature of the star by inverting the Stefan-Boltzmann law \citep{vanBelle1999} as $T_{\rm eff} = \left( \dfrac{4F_ {\rm bol}}{\sigma\theta_{\rm LD}^2 }\right)^{\frac{1}{4}}$,

where $\sigma$ is the Stefan–Boltzmann constant. The computed values of $T_{\rm eff}$, using the different LD prescriptions explained in \secref{vis-model}, are listed in \tabref{tab:interResults}. Because the STAGGER-based fit yields a larger $\theta_{LD}$, the derived $T_{\rm eff}$ is correspondingly lower by about 1~per cent. For the present analysis, we adopt $R$ and $T_{\rm eff}$ obtained using the direct fit pertaining to the four-parameter non-linear LD law, i.e. LD4 (Direct) of \tabref{tab:interResults}.

\begin{table*}
    \centering
    \renewcommand{\arraystretch}{1.2} 
    \setlength{\tabcolsep}{8pt}
    \begin{tabular}{ccccc}
    \hline
     Method & $u$ & $\theta_{\rm LD}$ & $R$ & $T_{\rm eff}$ \\
     & & (mas) & (R$_\odot$) & (K)\\
      \hline
       UD & 0 & $2.027\pm0.001$ & $8.23\pm0.01$ & $5196\pm33$\\
       LD4 (STAGGER) & - & $2.176^{+0.007}_{-0.037}$ & $8.83^{+0.07}_{-0.11}$ & $5015^{+60}_{-49}$ \\
       LD1 (STAGGER) & $0.621^{+0.019}_{-0.046}$ & $2.196^{+0.007}_{-0.016}$ & $8.90^{+0.07}_{-0.09}$ & $4992^{+54}_{-48}$\\
       LD4 (Direct) & - & $2.132^{+0.007}_{-0.005}$ & $8.65^{+0.10}_{-0.10}$ & $5066^{+47}_{-50}$\\
       LD1 (Direct) & $0.393^{+0.005}_{-0.005}$ & $2.125^{+0.002}_{-0.002}$ &
       $8.63^{+0.04}_{-0.04}$ & $5075^{+41}_{-41}$\\
      \hline
    \end{tabular}
    \caption{Angular diameters ($\theta_{\rm LD}$), physical radii ($R$) and temperatures ($T_{\rm eff}$) from different methods. UD denotes uniform-disc, LD1 a linear limb-darkening (LD) with coefficient $u$, and LD4 a four-parameter LD. The LD4 coefficients are not listed, as their numerical values are not directly informative (see \figref{fig:cornera}). The corresponding intensity profiles for LD4 are shown in \figref{fig:clv}.
    }.
    \label{tab:interResults}
\end{table*}

\begin{table*} 
    \centering
    \small  
    \renewcommand{\arraystretch}{1.2} 
    \begin{adjustbox}{max width=\textwidth}
    \begin{tabular}{ccccccccccc}
        \hline
        Filter & $U^a$  & $B^a$ & $V^a$ & $R^a$ & $G^b$ & $I^a$ & $J^a$ & $H^c$ & $K^a$ & $L^a$\\
            ($\lambda$ (nm)) & (360) & (450) & (555) & (670) & (673) & (870) & (1220) & (1630) & (2190) & (3500)\\
        \hline 
        mag & 5.47(5) & 4.73(5) & 3.76(5) & 3.13(5) & 3.54(5) & 2.66(5) & 2.21(5) & 1.82(18) & 1.64(5) & 1.58(9)\\
        Flux &
        \multirow{2}{*}{2.71(25)}&
        \multirow{2}{*}{8.11(75)}&
        \multirow{2}{*}{11.4(10)}&
        \multirow{2}{*}{12.2(11)}&
        \multirow{2}{*}{9.90(91)}&
        \multirow{2}{*}{9.72(90)}&
        \multirow{2}{*}{4.11(38)}&
        \multirow{2}{*}{2.12(68)}&
        \multirow{2}{*}{0.87(15)}&
        \multirow{2}{*}{0.17(3)}\\
        ($10^{-9}$ $\mathrm{erg\,s^{-1}\,cm^{-2}\,nm^{-1}}$)
        \\
        \hline
        \end{tabular}
    \end{adjustbox}
    
    \medskip
\ac{{\footnotesize
$^{a}$  from \citet{Ducati2002},
$^{b}$  from \citet{Gaia2022},
$^{c}$  from \citet{Cutri2003}}}

    \caption{Photometric magnitudes and fluxes of $\kappa$~Cyg from literature with wavelengths in nm for each filter.}
    \label{tab:stellar_mag}
\end{table*}

\section{Asteroseismology}

\subsection{Observations}
\label{astero-obs}

\begin{figure}
    \begin{subfigure}{0.48\textwidth}
       \includegraphics[width=\columnwidth]{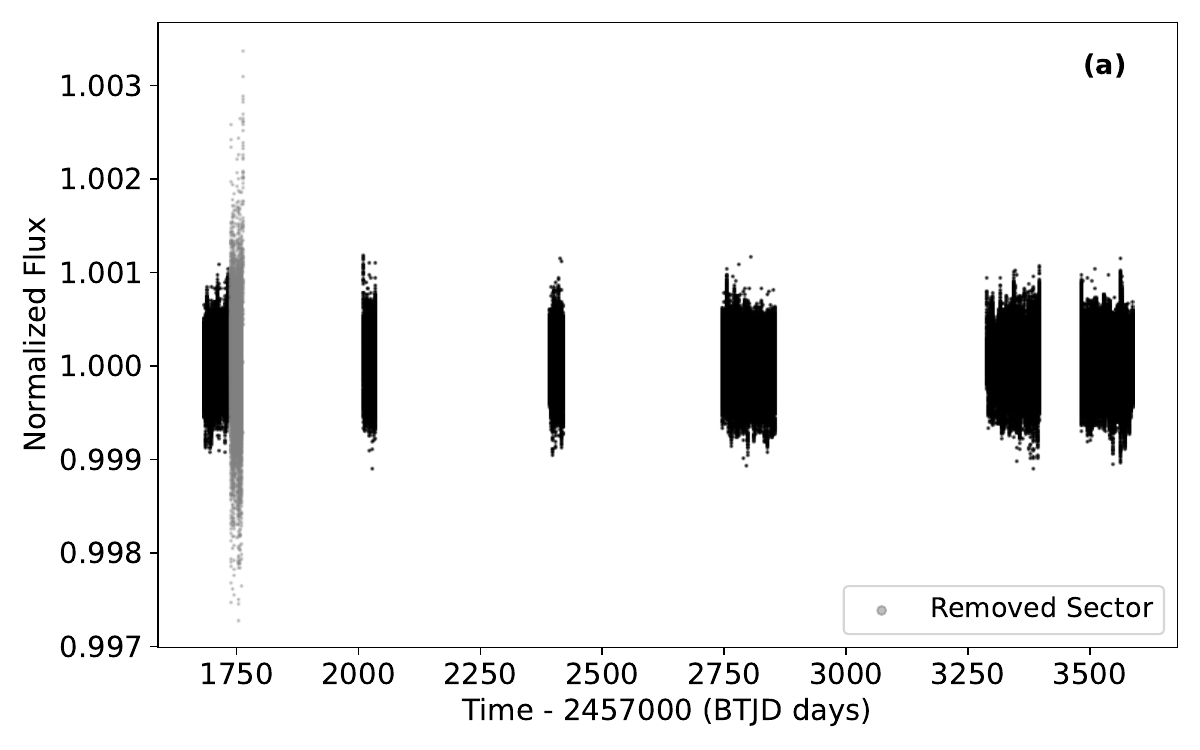}
       \phantomcaption
       \label{fig:lc}
    \end{subfigure}

    \begin{subfigure}{0.48\textwidth}
        \includegraphics[width=\columnwidth]{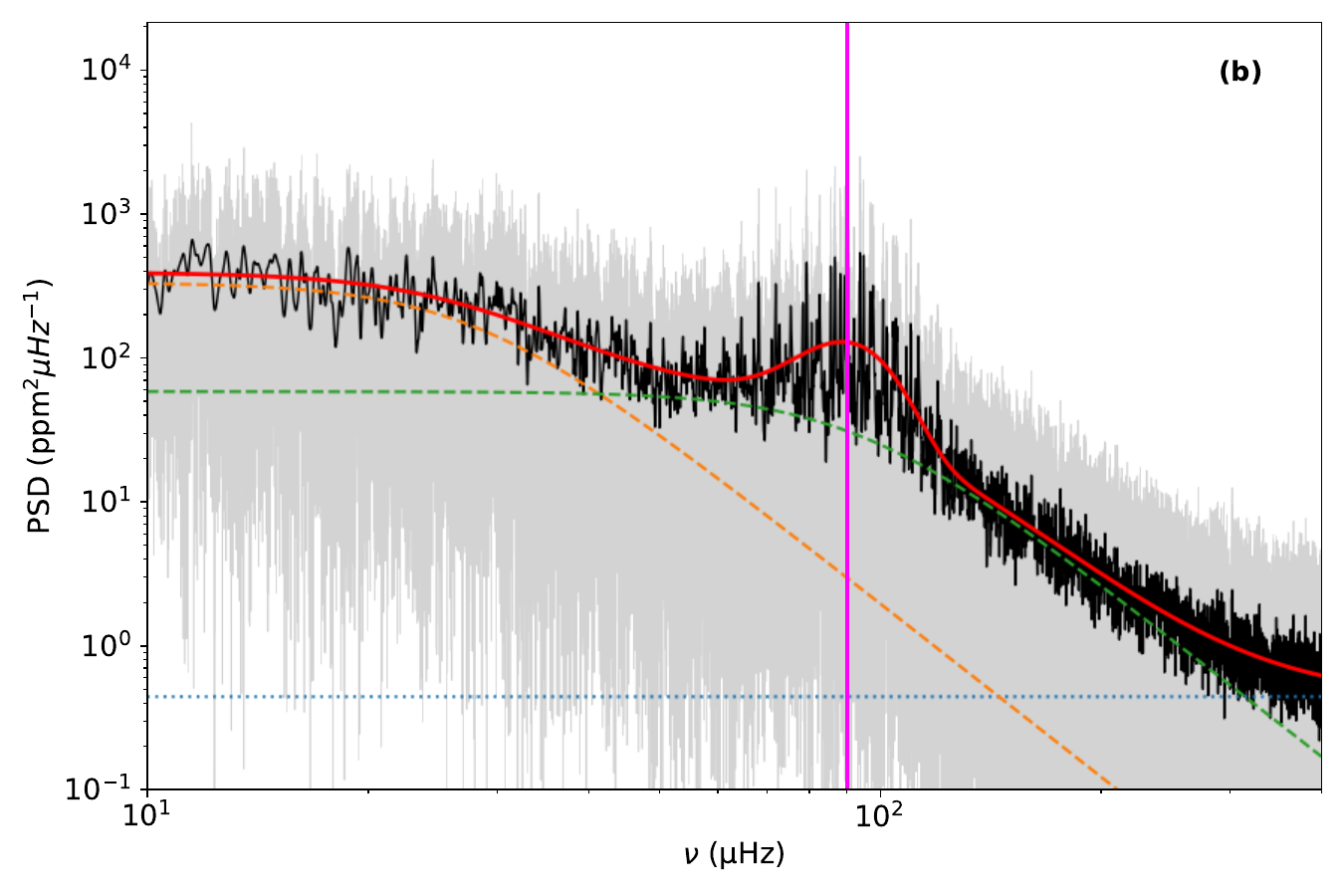}
        \phantomcaption
        \label{fig:ps}
    \end{subfigure}
   
   \caption{\textbf{(a)} The light curve of $\kappa$~Cyg with 16 sectors of \tess{} data, processed by the SPOC pipeline. Sector-16 \tess{} data (shown in grey) were omitted because there was a dead pixel column in the TPF. \textbf{(b)} The power spectral density (PSD) of $\kappa$~Cyg with the granulation background noise being modelled using two Harvey functions, shown in orange and green dashed lines. The horizontal dark-blue dotted line represents the white noise. $\nu_{\rm max}$ is shown in a vertical magenta line, and the total PSD fit is shown in a solid red line.}
\end{figure}

Our asteroseismic analysis relies on high-precision photometry obtained by the \textit{Transiting Exoplanet Survey Satellite} (\tess; \citealt{Ricker2015}). \tess{} observed $\kappa$~Cyg with a sampling cadence of 120~s across 17 sectors over six years. We used the light curve from the Science Performance Operation Center (SPOC) pipeline \citep{Jon2016}, which we downloaded from MAST using the lightkurve package \citep{LK2018}. Prior to analysis, the data from Sector 16 were removed since they were affected by a dead pixel column. This left us with a clean dataset spanning 16 sectors, shown in \figref{fig:lc}.

We calculated the power spectrum using the Lomb-Scargle method \citep{VanderPlas2018}, oversampled by a factor of 15 relative to the total observation time, and normalised to a power spectral density (PSD) following Parseval's theorem. We modelled the PSD as the sum of white noise from photon statistics ($P_{w}$), two super-Lorentzian Harvey-like components describing granulation (following \citealt{Kallinger2014}), and a Gaussian envelope representing the oscillation power excess. The full model is given by
\begin{equation}
P(\nu) = P_{w} + \sum_{i=1}^{2} \left[ \frac{a_i^2 / b_i}{1 + \left(\nu / b_i\right)^{4}} \right] + P_{g} \exp\left[-\frac{(\nu - \nu_{\rm max})^2}{2\sigma^2}\right],
\label{eq:background-model}
\end{equation}
where $a_i$ are the rms amplitudes, and $b_i$ are their corresponding characteristic frequencies ($\approx(2\pi\tau_i)^{-1}$, with $\tau_i$ being the characteristic granulation timescales). The Gaussian term has height $P_{g}$, central frequency $\nu_{\rm max}$, and width $\sigma$. All the components were fitted to the PSD simultaneously to avoid biases in the determination of $\nu_{\rm max}$. The resulting fit is shown in Fig.~\ref{fig:ps}. The oscillation power excess is well defined, exhibiting a clear Gaussian envelope with regularly spaced modes, allowing a reliable determination of the global asteroseismic parameters.

\subsection{Global Parameters}
\label{global-params}
The overall properties of solar-like oscillators are characterized by two global asteroseismic parameters:  the frequency of maximum power ($\nu_{\max}$) and the large frequency separation ($\Delta \nu$). 
The power excess of solar-like oscillations has a characteristic Gaussian envelope, the central frequency of which is denoted by $\nu_{\rm max}$. \citet{Brown1991} argued that $\nu_{\max}$ scales as the acoustic cut-off frequency ($\nu_{\rm ac}$) for solar-like oscillators, leading \citet{Kjeldsen1995} to posit the following scaling relation:
\begin{equation}
    \nu_{\max} \propto gT_{\rm eff}^{-0.5}.
    \label{eq:numax-nuac}
\end{equation}
 As mentioned in \secref{astero-obs}, we determined $\nu_{\max}$ by simultaneously fitting a Gaussian function along with the background noise to the PSD. We measured $\nu_{\rm max}=90.3\pm1.9$~$\mu$Hz. The quoted uncertainties were derived from the scatter observed when determining $\nu_{\max}$ for individual \tess{} visits across the analysed sectors (following \citealt{Sreenivas2024}). \ac{We measured an oscillation amplitude of $\sim25$~ppm per radial mode, following the method of \citep{kjeldsen2008b} with $c=3.04$ \citep{bedding2010, Yu2018}. This is as expected for a star with a $\nu_{\rm max}\sim90\ \mu$Hz (fig. 10 from \citealt{Yu2018}). Moreover, we can see no sign of flux dilution in the power spectral density of $\kappa$~Cyg: panel (a) of \figref{fig:echelles1} shows a similar height to the panels (a) and (e) of \figref{fig:echellesKepler}. This confirms our assertion from \secref{properties} that there is negligible photon count from the probable companion.}

The parameter $\Delta\nu$ is defined as the average frequency spacing between acoustic modes of the same angular degree $\ell$ and consecutive radial orders $n$ in the asymptotic regime ($n \gg \ell$). It is proportional to the inverse of the sound travel time across the stellar diameter and serves as a robust probe to the mean stellar density \citep{Ulrich1986}. This gives us the other asteroseismic scaling relation:
\begin{equation}
    \Delta\nu \propto\sqrt{\dfrac{M}{R^3}}.
    \label{eq:dnu-density}
\end{equation}
We measured $\Delta\nu=7.42\pm0.04~\mu$Hz for $\kappa$~Cyg using the autocorrelation of its power spectrum. Acoustic glitches can introduce systematic variations in $\Delta\nu$ \citep{Gough1990, Vrard2015, Verma2022} and so, to validate this measurement, we also fitted a straight line to the radial ($\ell=0$) mode frequencies as a function of radial order. Following the prescription of \citet{White2011}, we weighted the fit by a Gaussian envelope centred on $\nu_{\rm max}$, and determined $\Delta\nu = 7.42 \pm 0.03~\mu$Hz, emphasizing the robustness of $\Delta\nu$ determination.

Using the fundamental solar parameters from \citet{Huber2011} ($ \nu_{\max,\odot}=3090 \pm 30~\mu \text{Hz} $, $\Delta \nu_{\odot}=135.1 \pm 0.1~\mu \text{Hz}$), plus $T_{\mathrm{eff},\odot}=5772.0 \pm 0.8 ~ \text{K}$ \citep{Prsa2016}, we estimate the mass ($M$) and radius ($R$) of $\kappa$~Cyg by inverting the scaling relations \citep{Stello2008, Kallinger2010} as:

\begin{equation}
    \frac{R}{R_{\odot}} \approx \left(\frac{\nu_{\max}}{\nu_{\max,\odot}}\right) \left(\frac{\Delta\nu}{\Delta\nu_{\odot}f_{\Delta\nu}}\right)^{-2} \left(\frac{T_{\mathrm{eff}}}{T_{\mathrm{eff},\odot}}\right)^{0.5}  
    \label{eq:scalingRadius}
\end{equation}
and
\begin{equation}
    \frac{M}{M_{\odot}} \approx \left(\frac{\nu_{\max}}{\nu_{\max,\odot}}\right)^3 \left(\frac{\Delta\nu}{\Delta\nu_{\odot}f_{\Delta\nu}}\right)^{-4} \left(\frac{T_{\mathrm{eff}}}{T_{\mathrm{eff},\odot}}\right)^{1.5}.
    \label{eq:scalingMass}
\end{equation}
The correction factor $f_{\Delta\nu}$ (which accounts for deviations from \eqnref{eq:dnu-density}) is typically close to unity. It was estimated for RGB stars by \citet{Li2023} and, using a similar prescription, for CHeB stars by \citet{Schimak2026}. Using the latter, with $T_{\mathrm{eff}} = 5066^{+47}_{-50}$~K from \secref{fbol} and $\rm [Fe/H]=0.10\pm0.07$ from \secref{properties}, we obtained $f_{\Delta\nu}=0.995\pm0.002$. This gives $R = 8.96\pm0.22 ~ \rm R_{\odot}$ and $M = 2.19\pm0.16 ~ \rm M_{\odot}$. These values provide the initial constraints for the detailed stellar modelling of $\kappa$~Cyg discussed in \secref{model-desc}. \ac{We have not applied the correction term, $f_{\nu_{\rm max}}$ to $\nu_{\rm max}$ because it has been shown that $f_{\nu_{\rm max}}\sim1$ for solar metallicity stars \citep[e.g.,][]{Viani2017,Lundkvist2025}.}

\subsection{Individual Modes}
\label{kappa-modes}
The identification of individual oscillation mode frequencies ($ \nu_{n, \ell}$) is essential for detailed stellar modelling beyond scaling relations. To extract the mode frequencies, we used the background-divided power spectrum from \secref{astero-obs}. We obtained initial estimates of mode frequencies using a peak-finding algorithm implemented in \textsc{SciPy} \citep{Virtanen2020}, applied to a smoothed version of the PSD. The power spectrum of a stochastically excited, damped oscillation has a Lorentzian profile \citep{Anderson1990}, and we fitted this function to the background-divided PSD \citep{Garcia2018} within a $0.01\Delta\nu$ window centred on each detected peak:
\begin{equation}
    P(\nu)=\frac{H}{1+\left(\frac{2\left(\nu-\nu_0\right)}{\Gamma}\right)^2}.
    \label{eq:lorentzian-peaks}
\end{equation}
$P(\nu)$ is the background-divided power spectrum, $H$ is the mode height above the background (dimensionless), $\nu_0$ is the mode frequency, and $\Gamma$ is the full-width at half-maximum, related to the mode lifetime $\tau$ as $\Gamma = (\pi\tau)^{-1}$.

Uncertainties on the individual mode frequencies were estimated from the posterior distributions of a Markov Chain Monte Carlo (MCMC) sampler. We carried out the mode identification based on the asymptotic relation,
\begin{equation}
   \nu_{n,l} \simeq \Delta \nu\left(n+\frac{l}{2}+\epsilon_{p}(n, \ell)\right),
    \label{eq:asymptotic-relation}
\end{equation}
where $n$ and $\ell$ are the radial order and angular degree of the mode, respectively, and $\epsilon_{p}(n, \ell)$ is the phase offset. A complete list of oscillation mode frequencies detected in $\kappa$~Cyg is given in \tabref{tab:kapCygModes}.

For evolved stars like $\kappa$~Cyg, the oscillation modes have a mixed character, resulting from the coupling between pressure (p) modes in the envelope and gravity (g) modes in the core. In the asymptotic regime, while p-modes are equally spaced in frequency by $\Delta\nu$, the g-modes are equally spaced in period, defined by the g-mode period spacing for degree $\ell$ as $\Delta\Pi_\ell$. It is known that $\Delta\Pi_1$ is a diagnostic of the internal structure, particularly distinguishing between red giant branch (RGB) stars and core helium-burning (CHeB) clump stars \citep{bedding2011, Kallinger2012, mosser2012}. To focus our attention primarily on the stellar envelope, we decouple these mixed modes and isolate the pure p-mode characteristics (the $\pi$-modes) and the pure g-mode characteristics (the $\gamma$-modes) \citep{Ong2020, Ong2023}, we computed the stretched period ($\tau$) \citep{Mosser2015} and stretched frequency ($f$) (\citealt{Schimak2026}, \citealt{Liyg2024}, Y. Li et al, in prep). We calculated the Fourier transform of the stretched period power spectrum, and its maximum amplitude corresponds to $\Delta\Pi_1$. We obtained $\Delta\Pi_1=263.3\pm2.4$~s. The uncertainty was calculated using the formula for dense mixed-mode pattern from \citep{Vrard2016}. The high value of $\Delta\Pi_1$ identifies $\kappa$~Cyg as a CHeB star, and its high $\Delta\nu$ of $7.42\pm0.04\mu$Hz (see \secref{global-params}) confirms that it is a secondary red clump star \citep{bedding2011}.

\begin{figure*}
    \centering
    \includegraphics[width=\textwidth]{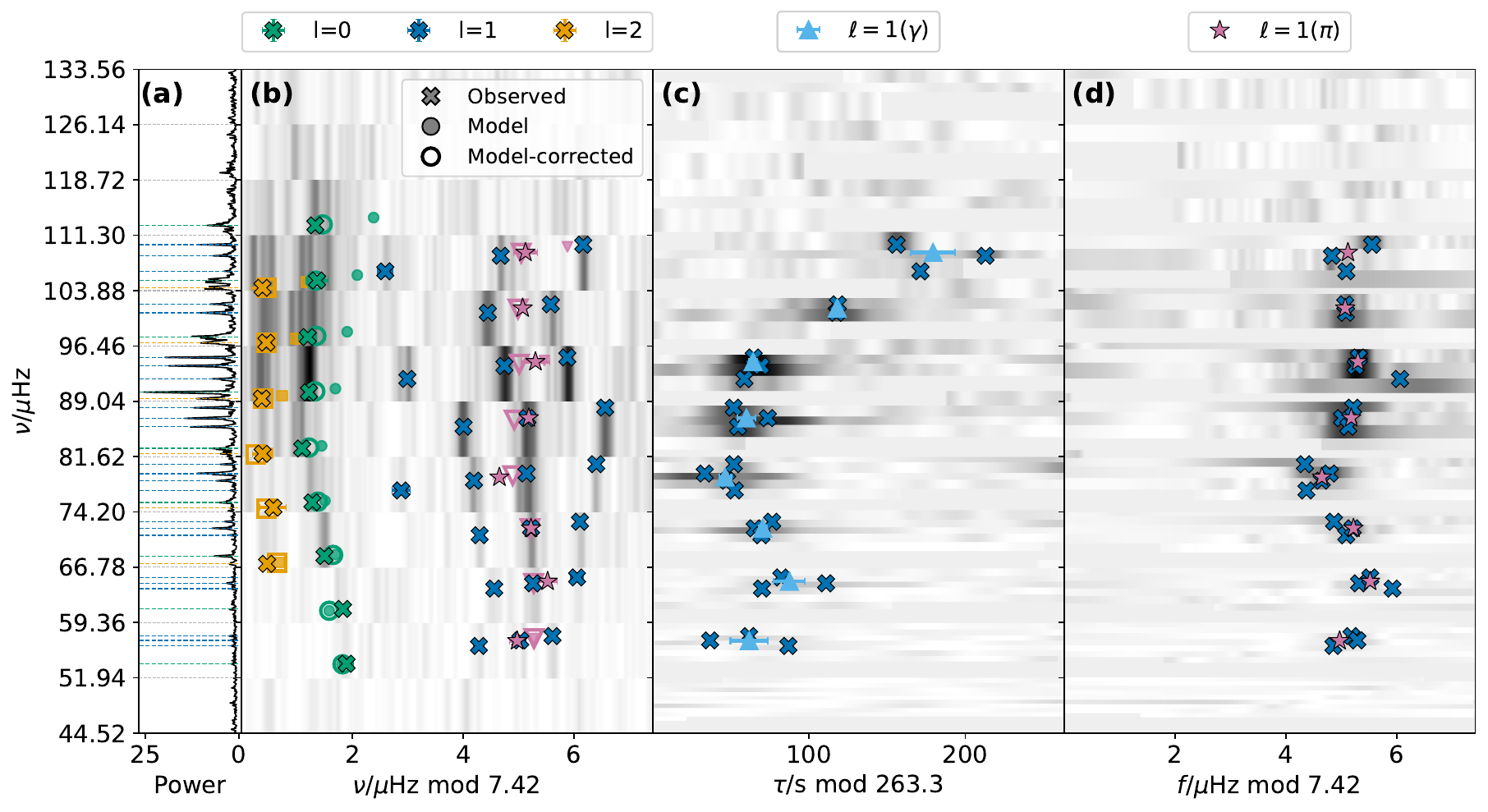}  
    \caption{The \echelle{} diagrams of $\kappa$~Cyg. \textbf{a:} Background-divided power spectrum with identified modes, shown in vertical dashed lines, green for radial, dark blue for mixed dipolar and orange for quadrupolar modes. \textbf{b:} Corresponding frequency \echelle{} diagram showing the observed modes with crosses and pure $l = 1$ $\pi$-modes in pink stars. Filled symbols are used for modelled frequencies \ac{(best from grid-OS)}, and the surface-corrected ones are shown with empty symbols (circle for $\ell=0$, inverted triangle for $\ell=1$ and square for $\ell=2$).     \textbf{c:} Corresponding stretched period \echelle{} diagram for the dipolar modes showing the pure $l$ = 1 $\gamma$-modes in light blue. \textbf{d:} Corresponding stretched frequency \echelle{} diagram for the dipolar modes.  We have masked out the radial and quadrupolar modes for better visualization in the stretched \echelle{} diagrams.}
    \label{fig:echelles1}
\end{figure*}

We determined the mode parameters (acoustic phase offset $\epsilon_p$, buoyancy phase offset $\epsilon_g$ and mode coupling strength $q$) by simultaneously aligning the $\gamma$-modes vertically in the stretched period \echelle{} diagram and the $\pi$-modes vertically in the stretched frequency \echelle{} diagram. The best-fit parameters correspond to minimal deviations of $\epsilon$ per acoustic radial order in both \echelle{} diagrams. We adopt the corresponding standard deviations as the uncertainties on respective $\pi$ and $\gamma$ modes. The resulting diagrams are shown in \figref{fig:echelles1}. We also note that there is a buoyancy ``glitch'', which causes curvature in the stretched period diagram, as seen in the third panel of \figref{fig:echelles1}. This is related to the sharp variations in the Brunt--V\"ais\"al\"a frequency profile in clump stars, owing to strong chemical gradients \citep{Cunha2019, Vrard2022, Matteuzzi2025}.

\begin{table}
\centering
\begin{tabular}{c c c c c}
\hline
$n_p$ & $\nu_{n,0}$ & $\nu_{n,1,\mathrm{mixed}}$ & $\nu_{n,1,\pi}$ & $\nu_{n,2}$ \\
 & ($\mu$Hz) & ($\mu$Hz) & ($\mu$Hz) & ($\mu$Hz) \\
 \hline
\noalign{\smallskip}
7 & \begin{tabular}[c]{@{}c@{}}53.833(127)\end{tabular} & \begin{tabular}[c]{@{}c@{}}56.221(98) \\ 56.967(98) \\ 57.548(60)\end{tabular} & 56.91(13) &  \\
\noalign{\smallskip}
8 & \begin{tabular}[c]{@{}c@{}}61.188(107)\end{tabular} & \begin{tabular}[c]{@{}c@{}}63.918(80) \\ 64.618(65) \\ 65.411(114)\end{tabular} & 64.88(18) & \begin{tabular}[c]{@{}c@{}}67.238(86)\end{tabular} \\
\noalign{\smallskip}
9 & \begin{tabular}[c]{@{}c@{}}68.272(29)\end{tabular} & \begin{tabular}[c]{@{}c@{}}71.073(71) \\ 72.001(59) \\ 72.884(88)\end{tabular} & 72.00(10) & \begin{tabular}[c]{@{}c@{}}74.771(238)\end{tabular} \\
\noalign{\smallskip}
10 & \begin{tabular}[c]{@{}c@{}}75.475(52)\end{tabular} & \begin{tabular}[c]{@{}c@{}}77.080(160) \\ 78.392(52) \\ 79.340(40) \\ 80.599(59)\end{tabular} & 78.85(11) & \begin{tabular}[c]{@{}c@{}}81.997(176)\end{tabular} \\
\noalign{\smallskip}
11 & \begin{tabular}[c]{@{}c@{}}82.711(42)\end{tabular} & \begin{tabular}[c]{@{}c@{}}85.624(33) \\ 86.776(31) \\ 88.179(31)\end{tabular} & 86.80(6) & \begin{tabular}[c]{@{}c@{}}89.403(32)\end{tabular} \\
\noalign{\smallskip}
12 & \begin{tabular}[c]{@{}c@{}}90.254(15)\end{tabular} & \begin{tabular}[c]{@{}c@{}}92.028(58) \\ 93.780(27) \\ 94.915(23)\end{tabular} & 94.34(26) & \begin{tabular}[c]{@{}c@{}}96.906(37)\end{tabular} \\
\noalign{\smallskip}
13 & \begin{tabular}[c]{@{}c@{}}97.653(164)\end{tabular} & \begin{tabular}[c]{@{}c@{}}100.901(44) \\ 102.037(69)\end{tabular} & 101.53(7) & \begin{tabular}[c]{@{}c@{}}104.255(201)\end{tabular} \\
\noalign{\smallskip}
14 & \begin{tabular}[c]{@{}c@{}}105.241(197)\end{tabular} & \begin{tabular}[c]{@{}c@{}}106.470(142) \\ 108.553(89) \\ 110.043(35)\end{tabular} & 109.00(22) &  \\
\noalign{\smallskip}
15 & \begin{tabular}[c]{@{}c@{}}112.628(51)\end{tabular} &  &  &  \\
\noalign{\smallskip}
\hline
\end{tabular}
\caption{Oscillation mode frequencies for $\kappa$~Cyg.}
\label{tab:kapCygModes}
\end{table}

\section{Modelling}
\subsection{Description of the models}
\label{model-desc}

To interpret the asteroseismic and interferometric observations of $\kappa$~Cyg, we computed two extensive grids of stellar evolutionary models using the open-source software, Modules for Experiments in Stellar Astrophysics  (\textsc{mesa}, version r24.03.1:\citealt{Paxton2011, Paxton2013, Paxton2015, Paxton2018, Paxton2019, Jermyn2023}). 

We evolved all evolutionary tracks from the pre-main sequence through to the CHeB phase with resolution settings of $\texttt{time\_delta\_coeff} = 0.5$, and 
$\texttt{mesh\_delta\_coeff} = 0.5$, 
corresponding to approximately a factor of two increase in both temporal and spatial resolutions relative to the default \textsc{MESA} settings. We terminated the models when the core He mass fraction dropped to $10^{-5}$, indicating evolution to the AGB phase. We used the \texttt{pp\_cno\_extras\_o18\_ne22.net} nuclear network to model the reactions and isotopes relevant to this advanced evolutionary state. To include the outer layers of the atmosphere in the models, we integrated the atmospheric boundary conditions out to an optical depth of $\tau = 10^{-3}$, employing the standard grey atmosphere Eddington $T-\tau$ relation \citep{Eddington1926}, and ensuring opacity consistent with the local temperature and pressure throughout the atmosphere. 

The treatment of convective boundaries in CHeB stars remains a key source of systematic uncertainty, affecting predictions of core size, mixing, and evolutionary timescales \citep[e.g.,][]{constantino2015, bossini2017, Paxton2018, Noll2024}. To explore the impact of these uncertainties, we computed two grids using different convective boundary-mixing prescriptions. The first grid (hereafter grid-PM) adopts MESA’s predictive mixing (PM) prescription \citep{Paxton2018}, while the second grid (grid-OS) uses the more commonly employed exponential overshooting (OS) prescription \citep[e.g.,][]{Paxton2011, Moravveji2016, Hernandez2019P}.  

Both PM and OS locate the formal convective boundary at the point where $y=\nabla_{\rm rad} - \nabla_{\rm ad}$ changes sign (Schwarzschild criterion; \citealt{Schwarz1958}). However, they extend the convective region beyond the formal limit in distinct ways to account for the momentum of convective motions near this boundary. 

PM checks how superadiabatic the first cell (the ``candidate cell'') on the notional radiative side is and mixes it so that it has the same composition as the adjacent cell on the convective side. Then it recomputes $y$, and if $y>0$ on both sides of the convective boundary, the next cell becomes the ``candidate cell''. This is repeated until the candidate cell has $y<0$ after the proposed mixing \citep{Paxton2018}. 

The convective OS is assumed to be diffusive \citep{Zhang2013}, and for the exponential prescription \citep{Herwig2000}, it extends the mixed region by an exponentially decreasing diffusion coefficient beyond the formal convective boundary, depending on the overshoot parameter, $f_{\rm ov}$ \citep{Paxton2011}.

For both grids, the initial set of parameters was distributed quasi-randomly across the parameter space using a Sobol sequence \citep{Sobol1967} to ensure efficient non-Cartesian exploration of the parameter space (grid-PM: $2^{12}=4096$ models; grid-OS: $2^{10}=1024$ models). The following parameters are standardized across both grids:

\begin{itemize}
    \item \textbf{Initial Mass ($M_{\rm ini}$):} We constrained the search to a mass range centred on the scaling relation mass from \secref{global-params}. We restricted the grid to $M_{\rm ini} \in [2.0, 2.8]\ \rm M_\odot$.
    
    \item \textbf{Metallicity ($Z_{\rm ini}$):} We varied the initial metallicity in the range $Z_{\rm ini} \in [0.015, 0.03]$, which corresponds to $\rm [Fe/H] \in [-0.13, 0.17]$, adopting the solar composition from \citet{Grevesse1998}. This covers the range expected from spectroscopic measurements, mentioned in \secref{properties}.

    \item \textbf{Helium Abundance ($Y_{\rm ini}$):} Since the helium abundance is difficult to constrain spectroscopically due to the absence of He absorption lines in the spectra, we varied $Y_{\rm ini}$ from 0.25 to 0.35. It covers a broader range than what the primordial He enrichment law mentioned in \citet{Li2018} provides. This was done since this enrichment law is often said to be an oversimplification \citep{Nsamba2021}.

    \item \textbf{Mixing Length ($\alpha_{\rm MLT}$):} Convection was treated using the standard Mixing Length Theory (MLT; \citealt{Bohm1958, Cox1968}) in which convective transport is approximated as a diffusive process with characteristic length-scale $\alpha_{\rm MLT} H_P$. Here, $H_P$ is the local pressure-scale height and $\alpha_{\rm MLT}$ is a free parameter that directly controls the efficacy of convective energy transport and strongly influences $R$ and $T_{\rm eff}$. $\alpha_{\rm MLT}$ is typically calibrated against the Sun, yielding $\alpha_{\rm MLT} \approx 1.92$ for a standard solar model \citep{Paxton2011, Joyce2018}, but observations of evolved stars indicate that non-solar values of $\alpha_{\rm MLT}$ are often required to reproduce their global properties \citep{Piau2011, Ball2017, Tayar2017, Li2018}. Hence, we allowed $\alpha_{\rm MLT}$ to vary over a wide range ($1.6-2.3$) in our model grid.
    
    \item \textbf{Mass Loss ($\eta$):} Mass loss near the tip of RGB affects the mass of the CHeB star, although less so for the more massive stars since their RGB lifetime is short and hence the integrated mass loss is low \citep{Girardi1999}. However, we included RGB mass loss using the \citet{Reimers1975} prescription to check whether our results are sensitive to plausible uncertainties in mass-loss efficiency. Following the cluster analysis by \citet{Miglio2012}, we varied the efficiency parameter over $\eta \in [0.1,0.3]$. 

    Next, we considered grid-specific parameters that can potentially influence our results:

   \item \textbf{\texttt{predictive\_superad\_thresh}} (for grid-PM only): The formal boundary of PM is defined using \texttt{predictive\_superad\_thresh} where the mixing stops if $\dfrac{\nabla_{\rm rad}}{\nabla_{\rm ad}}-1$ drops below this threshold instead of 0 (from Schwarzschild criterion; \citealt{Schwarz1958}). To minimize the occurrence of core breathing pulses and splitting of the convective core \citep{Paxton2018, Paxton2019, Ostrowski2021}, we varied \texttt{predictive\_superad\_thresh} from 0 to 0.025, covering the suggested value for CHeB stars \citep{Paxton2018}.

    \item \textbf{Overshooting ($f_{\rm ov}$)} (for grid-OS only):  We employed the commonly adopted exponential overshooting scheme, varying $f_{\rm ov}\in[0.0001-0.025]$ at the boundaries between radiative and convective zones in the core and envelope. \ac{This range is consistent with values employed in individual frequency modelling of evolved solar-like oscillators \citep[e.g.,][]{Hernandez2016, Claret2018, Schimak2026}. As noted by \citet{Viani2020}, we also found that increasing $f_{\rm ov}$ leads to larger modelled radii. Therefore, we did not extend $f_{\rm ov}$ to the unphysically large values sometimes required to match $\Delta\Pi_1$ alone \citep[e.g.,][]{constantino2015, Noll2024}, in order to remain consistent with the interferometric radius.}
\end{itemize}

\ac{For subsequent stellar modelling, we did not use the mixed-mode frequencies, as these strongly depend on the assumptions about core structure and convective boundary mixing, which govern g-mode properties and their coupling to p-modes. By contrast, acoustic waves, being largely confined to the outer envelope, are predominantly sensitive to envelope structure, and less so to the inner radiative regions. Along with the $\ell=0$ modes, we therefore fitted our stellar models using the pure $\pi$-mode prescription from \citet{Ong2020} for $\ell=1, 2$ modes.} We computed the corresponding theoretical adiabatic oscillation frequencies using the pulsation code \textsc{GYRE} (\citealt{Townsend2013}; version 7.1), as implemented within MESA’s \texttt{run\_star\_extras modules} \citep{Bellinger2022}. We searched for frequencies on a linear grid between 30 and 140 $\mu$Hz, covering the full observed oscillation envelope around $\nu_{\rm max}$. The scan used 200 trial frequencies for radial modes and 1000 trial frequencies for $\ell=1, 2$ modes.

\subsection{\texorpdfstring{Modelling of $\kappa$~Cyg}{Modelling of kappa Cyg}}
\label{kappa-modelling}
Standard 1D stellar structure models are known to inadequately capture the physical conditions in the near-surface layers of stars, characterized by low density and vigorous, non-adiabatic turbulent motions that are oversimplified in mixing-length theory \citep{Grigahcene2005}. In addition, surface magnetism and non-adiabatic frequency shifts from radiative and convective damping also act predominantly in these layers. These deficiencies introduce a systematic offset in modelled oscillation frequencies, the so-called ``surface effect", which can significantly bias the inference of fundamental stellar parameters if left uncorrected.

To mitigate this systematic deviation, we employed the two-term parametric surface correction proposed by \citet{Ball2014}. Unlike the former empirical power-laws \citep{Kjeldsen2008}, this correction is derived from the theoretical work of \citet{Gough1990} and scales inversely with the normalised mode inertia $I_{n,l}$, which we obtained from \textsc{GYRE}.

We calculated the surface correction term, $\delta\nu_{\rm surf}$, for each model frequency, $\nu_{\rm mod}$,  as
\begin{equation}
     \frac{\delta\nu_{\rm surf}}{\nu_{\rm ac}} = \frac{1}{I_{n,l}} \left[ a_{-1} \left( \frac{\nu_{\rm mod}}{\nu_{\rm ac}} \right)^{-1} + a_3 \left( \frac{\nu_{\rm mod}}{\nu_{\rm ac}} \right)^{3} \right],
    \label{eq:BG14}
\end{equation}
where $\nu_{\rm ac}$ is the acoustic cutoff frequency, and $a_{-1}$ and $a_3$ are dimensionless coefficients determined by fitting the corresponding observed mode frequencies, $\nu_{\rm obs}$. For every model in our grid, we determined $a_{-1}$ and $a_3$ via a linear least-squares fit to the differences between $\nu_{\rm mod}$ and $\nu_{\rm obs}$. We then assessed the goodness-of-fit for each model using a reduced $\chi_{\rm red}^2$ metric defined as
\begin{equation}
    \chi^2_{\rm red} =  \frac{1}{N-2}\sum_{i=1}^{N} \frac{(\nu_{{\rm obs}, i} - \nu_{{\rm mod}, i} - \delta\nu_{{\rm surf}, i})^2}{\sigma_i^2},
    \label{eq:asteroLikely}
\end{equation}
where $N$ is the number of fitted modes, and $\sigma_i$ are the observational uncertainties on the individual mode frequencies (see \tabref{tab:kapCygModes}). We further constrained the sign of surface correction so that our analysis retained only the models in which the radial mode frequencies follow the expected trend, similar to solar models (i.e., $\delta\nu_{\rm surf}\leq0$, and its magnitude increases with frequency). We imposed a constraint on $T_{\rm eff}$ from \secref{fbol} and on no other parameter to test the asteroseismic performance of our two convective boundary mixing prescriptions. The likelihood for each model was calculated as $\mathcal{L}=\exp{(-(\chi^2_{\rm red}+\chi^2_{T_{\rm eff}})/2)}$.

\begin{figure*}
    \centering
    \includegraphics[width=\linewidth]{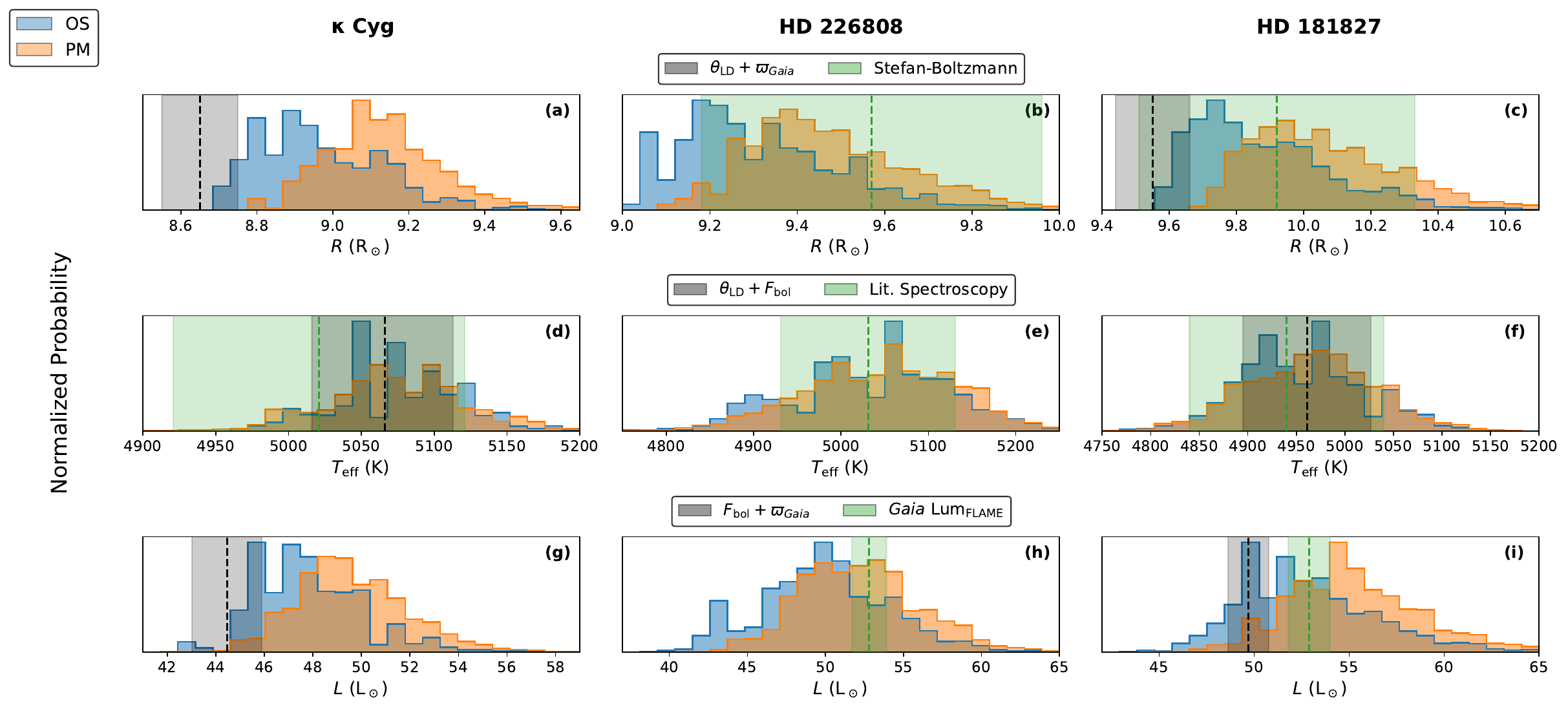}
    \caption{Overall modelling results: blue is for overshooting (OS), and orange is for predictive mixing (PM). \textbf{(a-c):} radius distributions. The grey patch is for PAVO radius with \gaia{} parallax, and the green patch is for Stefan-Boltzmann radius using \gaia{} $\rm Lum_{FLAME}$ and spectroscopic $T_{\rm eff}$. \textbf{(d-f):} $T_{\rm eff}$ distributions. The grey patch shows $T_{\rm eff}$ using bolometry and PAVO angular diameter. The green patch is for spectroscopic $T_{\rm eff}$. \textbf{(g-i):} luminosity distributions. The grey patch is $L$ from bolometry and \gaia{} parallax, and the green patch is from \gaia{} $\rm Lum_{FLAME}$.}
    \label{fig:histAllStars}
\end{figure*}

The weighted histograms of the stellar parameters for $\kappa$ Cyg derived from this process are presented in panels (a), (d) and (g) of \figref{fig:histAllStars}. The observed interferometric radius, shown using a grey-shaded patch, shows a significant discrepancy: the modelled radius distributions for both prescriptions of convective boundary mixing are overestimated. The best-fitting models from grid-OS, while marginally closer to the interferometric value, are still systematically higher by about $3.2$~per cent. The grid-PM shows a distinct lack of acceptable models capable of reproducing the required radius. \ac{We also note that the radius from grid-OS is very close to that from the scaling relations (from \secref{global-params}), whereas the radius from grid-PM shows a noticeable difference. This highlights a difference in the underlying physical assumptions of the models used to obtain $f_{\Delta\nu}$ and models in grid-PM}. The median $T_{\mathrm{eff}}$ values from both grids agree with the  $T_{\mathrm{eff}}$ from \secref{fbol} within $1\ \sigma$. Hence, the luminosity predicted by the models is also higher than what we calculated in \secref{fbol}. The modelled values for $\kappa$~Cyg using both models are reported in \tabref{tab:modelResults} and the corner plot corresponding to grid-OS is shown in \figref{fig:corner}. To further constrain our system, we considered including additional parameters in the likelihood function. \ac{Including $\nu_{\rm max}$ did not alter the results. As shown in \tabref{tab:modelResults} and \figref{fig:corner}, $\nu_{\rm max}$ is accurately reproduced even without additional constraints.} We also explored the use of $\Delta\Pi$ and $R$ as additional constraints; however, the grids did not contain sufficient models to incorporate them meaningfully. The modelled $\Delta\Pi_1$ is smaller than the observed one, in line with previous studies \citep[e.g.,][]{constantino2015, Noll2025}. We tested including a third term, linearly dependent on frequency in \eqnref{eq:BG14}. However, this had little impact on our results, indicating that the primary source of the radius discrepancy is unlikely to be the difference in density perturbations confined to the near-surface layers of the models and the actual star.

\begin{table*}
\centering
\small
\renewcommand{\arraystretch}{1.2}

\begin{adjustbox}{max width=\textwidth}
\begin{tabular}{lccccccccc}
\hline
Source  & $M$ & $R$ & $T_{\rm eff}$ & $L$ & Age & $\nu_{\rm max}$ & $\Delta\Pi_1$ & $\log g$ & [Fe/H] \\
       & ($M_\odot$) & ($R_\odot$) & (K) & ($L_\odot$) & (Gyr) & ($\mu$Hz) & (s) & & \\
\hline

\multicolumn{10}{c}{{HD~181276 ($\kappa$~Cyg)}} \\
\tess~+~PAVO & --- & $8.65_{-0.10}^{+0.10}$ & $5066_{-47}^{+50}$ & 
$44.46_{-1.43}^{+1.43}$ & --- & $90.3^{+1.9}_{-1.9}$ & $263.3^{+2.4}_{-2.4}$ &  $3.02^{+0.08}_{-0.08}$\ $^a$ & $0.10^{+0.07}_{-0.07}$\ $^a$  \\
Grid (PM) & $2.32_{-0.08}^{+0.11}$ & $9.12_{-0.13}^{+0.15}$ & $5073_{-49}^{+42}$ & $49.41_{-2.23}^{+2.32}$ & $0.68_{-0.10}^{+0.12}$ & $92.20_{-1.08}^{+1.40}$ & $206_{-34}^{+45}$ & $2.884_{-0.003}^{+0.006}$ & $0.04_{-0.11}^{+0.09}$\\
Grid (OS) & $2.19_{-0.07}^{+0.14}$ & $8.93_{-0.13}^{+0.21}$ & $5077_{-41}^{+40}$ & $47.55_{-1.89}^{+2.27}$ & $0.88_{-0.15}^{+0.11}$ & $90.79_{-0.79}^{+1.56}$ & $208_{-39}^{+25}$ & $2.878_{-0.004}^{+0.007}$ & $0.06_{-0.13}^{+0.10}$\\
\addlinespace

\multicolumn{10}{c}{{HD~226808 (KIC~5307747)}} \\
\kepler~+~\gaia & --- & $9.57_{-0.39}^{+0.39}$ & $5031_{-100}^{+100}$ & $52.81_{-1.10}^{+1.10}$ & --- & $83.4^{+2.3}_{-2.3}$ & $290.0^{+1.5}_{-1.5}$ & $2.77_{-0.08}^{+0.08}$\ $^b$ & $0.01_{-0.07}^{+0.07}$\ $^b$\\
Grid (PM) & $2.26_{-0.09}^{+0.14}$ & $9.45_{-0.15}^{+0.22}$ & $5052_{-90}^{+82}$ & $52.09_{-3.67}^{+3.98}$ & $0.70_{-0.13}^{+0.12}$ & $84.02_{-1.48}^{+1.72}$ & $202_{-41}^{+68}$ & $2.843_{-0.005}^{+0.007}$ & $0.04_{-0.11}^{+0.09}$\\
Grid (OS) & $2.16_{-0.09}^{+0.14}$ & $9.28_{-0.14}^{+0.24}$ & $5048_{-113}^{+74}$ & $49.78_{-3.74}^{+4.48}$ & $0.89_{-0.18}^{+0.21}$ & $82.82_{-1.41}^{+1.85}$ & $202_{-36}^{+44}$ & $2.836_{-0.006}^{+0.008}$ & $0.06_{-0.09}^{+0.08}$ \\
\addlinespace

\multicolumn{10}{c}{{HD~181827 (KIC~8813946)}} \\
\kepler~+~PAVO & --- & $9.55_{-0.11}^{+0.11}$ & $4961_{-66}^{+66}$ & $49.70_{-1.07}^{+1.07}$ & --- & $73.1^{+2.3}_{-2.3}$ & $295.0_{-2.0}^{+2.0}$ & $2.81_{-0.08}^{+0.08}$\ $^c$ & $0.14_{-0.07}^{+0.07}$\ $^c$\\
Grid (PM) & $2.26_{-0.10}^{+0.18}$ & $10.03_{-0.18}^{+0.27}$ & $4968_{-72}^{+65}$ & $54.84_{-2.90}^{+3.77}$ & $0.69_{-0.15}^{+0.12}$ & $75.18_{-1.24}^{+1.84}$ & $219_{-35}^{+52}$ & $2.790_{-0.005}^{+0.010}$ & $0.06_{-0.09}^{+0.08}$ \\
Grid (OS) & $2.14_{-0.08}^{+0.16}$ & $9.82_{-0.14}^{+0.27}$ & $4948_{-50}^{+67}$ & $52.41_{-2.99}^{+4.13}$ & $0.88_{-0.18}^{+0.25}$ & $74.02_{-1.04}^{+1.60}$ & $225_{-36}^{+31}$ & $2.784_{-0.006}^{+0.009}$ & $0.06_{-0.09}^{+0.08}$ \\
\hline
\end{tabular}
\end{adjustbox}

\medskip
{\footnotesize
$^{a}$  from \citet{deka2018},
$^{b}$  from \citet{Takeda2015},
$^{c}$  from \citet{Thygesen2012}}
\caption{Comparison of observations with asteroseismic modelling results for key stellar parameters of the three targets, obtained using the predictive mixing (PM) and overshooting (OS) prescriptions.}
\label{tab:modelResults}
\end{table*}

\subsection{Comparison with two \kepler{} analogues}
\label{kepler-modelling}
Our difficulties in finding a model of $\kappa$~Cyg that is consistent with its interferometric radius may indicate either that $\kappa$~Cyg is anomalous in some fashion, or that secondary clump stars in general are systematically poorly described by stellar models of the kind that we have used. To investigate, we extended our analysis to two bright secondary clump stars observed by the \kepler{} mission \citep{Borucki2010}: HD~226808 ($V=7.19$) and HD~181827 ($V=8.67$), the second of which also has an interferometric radius. For both stars, we applied a methodology identical to $\kappa$~Cyg for mode extraction and grid modelling. The independent measurements used for comparison with the asteroseismic models can be found in \tabref{tab:modelResults}.

\subsubsection{The case of HD~226808}
\begin{figure*}
    \begin{subfigure}{\textwidth}
       \includegraphics[width=\columnwidth]{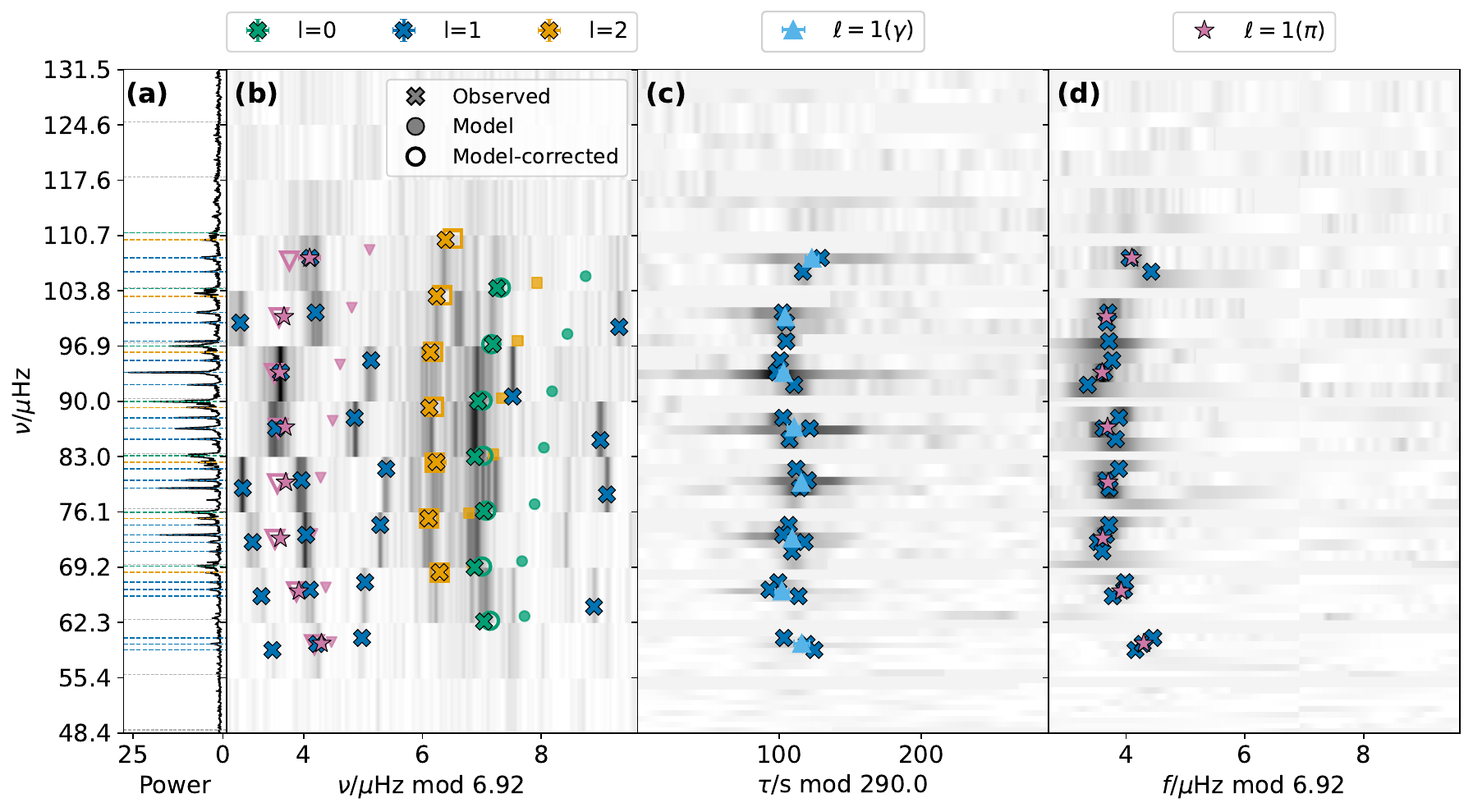}
       \phantomcaption
    \end{subfigure}

      \begin{subfigure}{\textwidth}
       \includegraphics[width=\columnwidth]{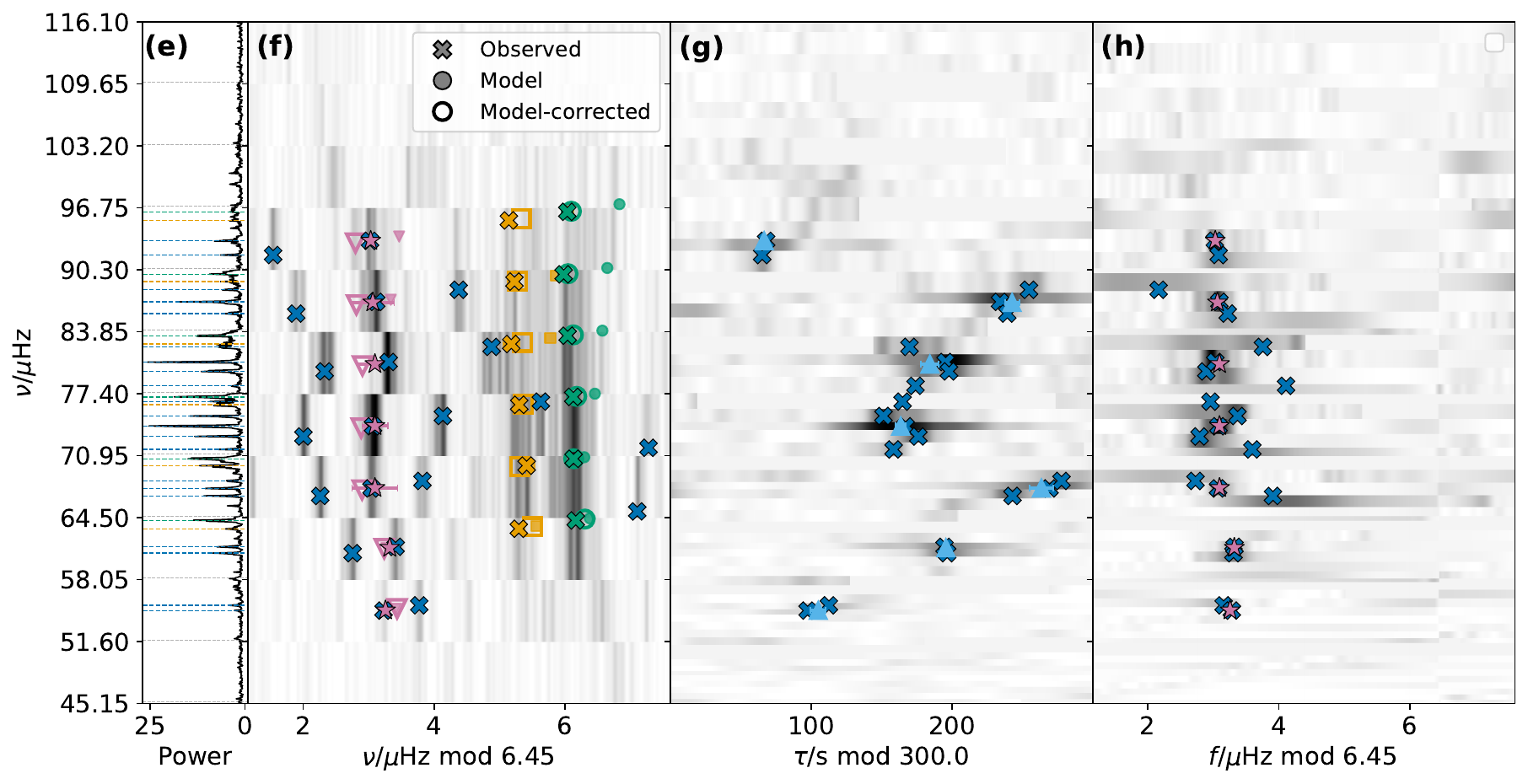}
       \phantomcaption
    \end{subfigure}

    \caption{Same as \figref{fig:echelles1} but for the \kepler{} stars (\ac{panels a-d for} HD~226808 on top and \ac{panels e-h for} HD~181827 at bottom). Note that the \echelle{} diagrams have been horizontally shifted for better visualization of the respective ridges.}
    \label{fig:echellesKepler}
\end{figure*}

\begin{table}
\centering
\begin{tabular}{c c c c c}
\hline
$n_p$ & $\nu_{n,0}$ & $\nu_{n,1,\mathrm{mixed}}$ & $\nu_{n,1,\pi}$ & $\nu_{n,2}$ \\
 & ($\mu$Hz) & ($\mu$Hz) & ($\mu$Hz) & ($\mu$Hz) \\
 \hline
\noalign{\smallskip}
8 &  & \begin{tabular}[c]{@{}c@{}}58.834(69) \\ 59.587(49) \\ 60.343(56)\end{tabular} & 59.66(8) &  \\
\noalign{\smallskip}
9 &  & \begin{tabular}[c]{@{}c@{}}65.570(80) \\ 66.386(40) \\ 67.314(34)\end{tabular} & 66.20(8) & \begin{tabular}[c]{@{}c@{}}68.569(45)\end{tabular} \\
\noalign{\smallskip}
10 & \begin{tabular}[c]{@{}c@{}}69.315(82)\end{tabular} & \begin{tabular}[c]{@{}c@{}}71.167(83) \\ 72.342(64) \\ 73.242(19) \\ 74.489(34)\end{tabular} & 72.81(8) & \begin{tabular}[c]{@{}c@{}}75.303(88)\end{tabular} \\
\noalign{\smallskip}
11 & \begin{tabular}[c]{@{}c@{}}76.078(45)\end{tabular} & \begin{tabular}[c]{@{}c@{}}79.096(30) \\ 80.085(25) \\ 81.509(30)\end{tabular} & 79.82(3) & \begin{tabular}[c]{@{}c@{}}82.356(161)\end{tabular} \\
\noalign{\smallskip}
12 & \begin{tabular}[c]{@{}c@{}}83.153(47)\end{tabular} & \begin{tabular}[c]{@{}c@{}}85.225(27) \\ 86.577(37) \\ 87.898(32)\end{tabular} & 86.73(7) & \begin{tabular}[c]{@{}c@{}}89.155(90)\end{tabular} \\
\noalign{\smallskip}
13 & \begin{tabular}[c]{@{}c@{}}89.922(10)\end{tabular} & \begin{tabular}[c]{@{}c@{}}92.035(23) \\ 93.578(19) \\ 95.096(38)\end{tabular} & 93.56(11) & \begin{tabular}[c]{@{}c@{}}96.094(130)\end{tabular} \\
\noalign{\smallskip}
14 & \begin{tabular}[c]{@{}c@{}}96.897(27)\end{tabular} & \begin{tabular}[c]{@{}c@{}}97.479(26) \\ 99.813(54) \\ 101.082(55)\end{tabular} & 100.55(15) & \begin{tabular}[c]{@{}c@{}}103.118(85)\end{tabular} \\
\noalign{\smallskip}
15 & \begin{tabular}[c]{@{}c@{}}104.065(64)\end{tabular} & \begin{tabular}[c]{@{}c@{}}106.191(33) \\ 107.913(67)\end{tabular} & 107.90(21) & \begin{tabular}[c]{@{}c@{}}110.189(54)\end{tabular} \\
\noalign{\smallskip}
16 & \begin{tabular}[c]{@{}c@{}}111.056(110)\end{tabular} &  &  &  \\
\noalign{\smallskip}
\hline
\end{tabular}
\caption{Oscillation mode frequencies for HD~226808.}
\label{tab:226808Modes}
\end{table}

HD~226808 (KIC~5307747) is one of the brightest \kepler{} secondary clump stars studied by \citet{Mosser2014}. The star has been previously modelled by \citet{Moura2020}, but the modes were misidentified ($\ell=0$ and $\ell=1$ ridges were swapped). We prepared its background-divided power spectrum and identified modes using stretched \echelle{} diagrams for $\pi$ and $\gamma$ modes, shown in panels (a)-(d) of \figref{fig:echellesKepler}. We do not see any curvature in the stretched period \echelle{} diagram. We expect this if the star is very early in its CHeB phase and has no sharp density gradient. The full list of its mode frequencies is presented in \tabref{tab:226808Modes}. Since high-fidelity interferometry is unavailable for HD~226808, we compared our derived luminosity with that reported in the \gaia{} DR3 catalogue using the FLAME module \citep{Creevey2022}. Contrary to what we observed for $\kappa$~Cyg, both model grids predict the \gaia{} luminosity quite accurately. Also, we used the spectroscopic $T_{\rm eff}$ reported from \citet{Takeda2015} and calculated its radius using the Stefan-Boltzmann law. The distributions of the modelled parameters can be seen in the panels (b), (e) and (h) of \figref{fig:histAllStars}. Since an interferometric radius is not available, a direct comparison and a firm conclusion cannot be made. 

\subsubsection{The case of HD~181827}

\begin{table}
\centering
\begin{tabular}{c c c c c}
\hline
$n_p$ & $\nu_{n,0}$ & $\nu_{n,1,\mathrm{mixed}}$ & $\nu_{n,1,\pi}$ & $\nu_{n,2}$ \\
 & ($\mu$Hz) & ($\mu$Hz) & ($\mu$Hz) & ($\mu$Hz) \\
 \hline
\noalign{\smallskip}
7 &  & \begin{tabular}[c]{@{}c@{}}54.828(57) \\ 55.373(37)\end{tabular} & 54.86(6) &  \\
\noalign{\smallskip}
8 &  & \begin{tabular}[c]{@{}c@{}}60.806(28) \\ 61.462(25)\end{tabular} & 61.37(3) & \begin{tabular}[c]{@{}c@{}}63.343(56)\end{tabular} \\
\noalign{\smallskip}
9 & \begin{tabular}[c]{@{}c@{}}64.213(64)\end{tabular} & \begin{tabular}[c]{@{}c@{}}66.761(26) \\ 67.544(17) \\ 68.322(58)\end{tabular} & 67.60(25) & \begin{tabular}[c]{@{}c@{}}69.913(41)\end{tabular} \\
\noalign{\smallskip}
10 & \begin{tabular}[c]{@{}c@{}}70.630(64)\end{tabular} & \begin{tabular}[c]{@{}c@{}}71.602(46) \\ 72.953(19) \\ 74.017(11) \\ 75.085(19) \\ 76.580(52)\end{tabular} & 74.05(18) & \begin{tabular}[c]{@{}c@{}}76.250(72)\end{tabular} \\
\noalign{\smallskip}
11 & \begin{tabular}[c]{@{}c@{}}77.075(16)\end{tabular} & \begin{tabular}[c]{@{}c@{}}78.225(71) \\ 79.727(25) \\ 80.706(17) \\ 82.279(64)\end{tabular} & 80.50(21) & \begin{tabular}[c]{@{}c@{}}82.578(130)\end{tabular} \\
\noalign{\smallskip}
12 & \begin{tabular}[c]{@{}c@{}}83.439(56)\end{tabular} & \begin{tabular}[c]{@{}c@{}}85.744(42) \\ 86.965(26) \\ 88.226(80)\end{tabular} & 86.91(15) & \begin{tabular}[c]{@{}c@{}}89.077(50)\end{tabular} \\
\noalign{\smallskip}
13 & \begin{tabular}[c]{@{}c@{}}89.828(51)\end{tabular} & \begin{tabular}[c]{@{}c@{}}91.841(39) \\ 93.321(52)\end{tabular} & 93.33(5) & \begin{tabular}[c]{@{}c@{}}95.441(67)\end{tabular} \\
\noalign{\smallskip}
14 & \begin{tabular}[c]{@{}c@{}}96.334(92)\end{tabular} &  &  &  \\
\noalign{\smallskip}
\hline
\end{tabular}
\caption{Oscillation mode frequencies for HD~181827.}
\label{tab:181827Modes}
\end{table}

HD~181827 (KIC~8813946) is another secondary-clump star in the \kepler{} field \citep{Mosser2014}. Its global asteroseismic properties and limb-darkened angular diameter, $\theta_{\rm LD}$, were previously measured by \citet{Huber2012}. As noted by \citet{Huber2012}, there were no \ac{visibility} data beyond the first null from PAVO interferometry for this star. Hence, we cannot employ our prescription to use the more realistic non-linear LD parametrization. Instead, we use the linear prescription. Motivated by our results for $\kappa$~Cyg, which suggest that atmospheric models may favour higher limb-darkening coefficients, we imposed a weak Gaussian prior on the linear limb-darkening coefficient ($u$) centred at 0.4 while fitting the PAVO data for $\theta_{\rm LD}$ and $u$. Our $\theta_{\rm LD}$ agreed with the reported interferometric value within $0.5$~per cent. 

 The \kepler{} background-divided power spectrum and \echelle{} diagrams for this star are shown in panels (e)-(h) of \figref{fig:echellesKepler}, and the complete set of its observed frequencies is listed in \tabref{tab:181827Modes}. As in $\kappa$~Cyg, we can notice a buoyancy ``glitch'' in the stretched period \echelle{} diagram of HD~181827.
 
 Using the updated interferometric radius derived from the \gaia{} parallax and the bolometric flux and temperature from PHOENIX models \citep{Allard2011, Allard2012}, we compared our asteroseismic modelling results (panels (c), (f) and (i) of \figref{fig:histAllStars}) against these direct constraints. The modelling results for HD~181827 echo the discrepancy encountered with $\kappa$~Cyg: the model-derived radii are systematically greater than the measured interferometric radius by 3.2~per cent. Hence, the luminosities derived from the model grids are higher than our bolometric calculation. However, the \gaia{} FLAME luminosity agrees quite well with our model grids. Also, the spectroscopic $T_{\rm eff}$ value from \citet{Thygesen2012} matches with what we calculated from bolometry. So, the Stefan-Boltzmann radius for the star, calculated using the \gaia{} FLAME luminosity, is significantly larger than PAVO $\theta_{\rm LD}$ and hence much closer to the asteroseismically modelled radius.

Although no independent constraints from interferometry, bolometry, or spectroscopy were imposed on our models, \figref{fig:histAllStars} reveals a clear lack of acceptable models from either grid that reproduce the interferometric radii of $\kappa$~Cyg and HD~181827.

\subsection{Analysis of $\epsilon_p$ differences}
\begin{figure}
    \centering
    \includegraphics[width=0.48\textwidth]{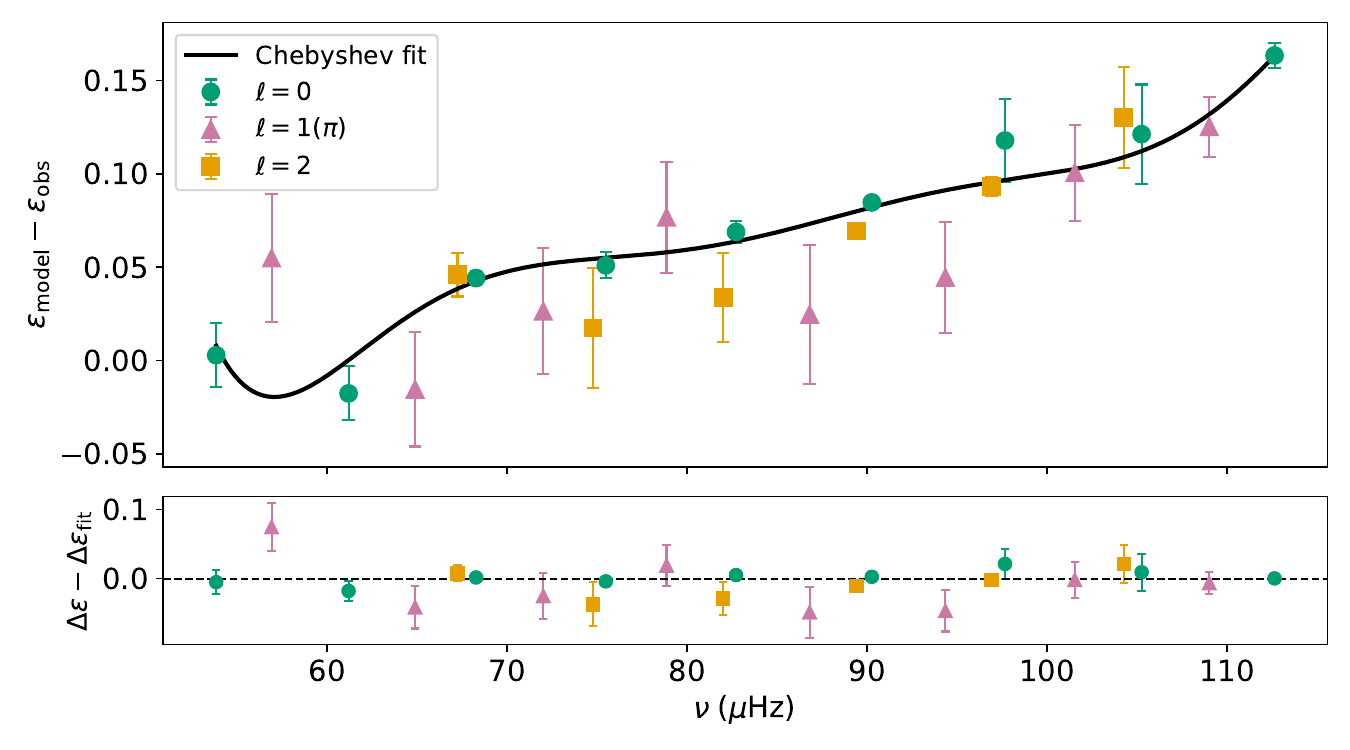}  
    \caption{Differences in $\epsilon_p$ \ac{for $\kappa$~Cyg} between the modelled (best from grid-OS) and observed frequencies along with the residuals from the best $\ell$-independent Chebyshev polynomial fit. \ac{The frequencies correspond to those shown in panel (b) of \figref{fig:echelles1}.}}
    \label{fig:epsModObs}
\end{figure}

Following \citet{Roxburgh2016} and \citet{Ong2021}, we looked at the differences in $\epsilon_p$ (from \eqnref{eq:asymptotic-relation}) between the best model and the observed frequencies to infer about the respective internal structures. \ac{For this analysis, we used the measured $\ell=0$ and $\ell= 2$ modes, and the inferred $\ell=1\ \pi$ modes.} If the two models share the same interior structure, the $\epsilon_p$ differences reduce to a function of $\nu$ alone. In contrast, if their internal structures differ, these differences depend on both $\ell$ and $\nu$. 

We first fitted an $\ell$-independent Chebyshev polynomial in frequency to the $\epsilon_p$ differences, representing the null hypothesis (H0) of no $\ell$-dependence (see \figref{fig:epsModObs}). To test for potential $\ell$-dependent structure (H1), we performed a nested-model comparison, which yields $\Delta\chi^2 = 8.85$ ($p = 0.012$), an F-test $p = 0.079$, $\Delta$AIC = 4.85, and $\Delta$BIC = 2.58, corresponding to weak-to-moderate evidence for $\ell$-dependence on the Jeffreys scale \citep{Kass1995}. The $\ell=1, 2$ modes show a systematic offset from the $\ell$-independent fit that is twice in magnitude to the overall weighted scatter. So, we note that the residuals reveal a subtle $\ell$-dependent pattern, likely driven by differences between the internal structure of the models and the star, particularly near the core boundary. Similar analyses using the corresponding best models for the two \kepler{} stars show comparable $\ell$-dependent signatures.

\begin{figure*}
    \centering
    \begin{subfigure}{0.48\linewidth}
        \centering
        \includegraphics[width=\linewidth]{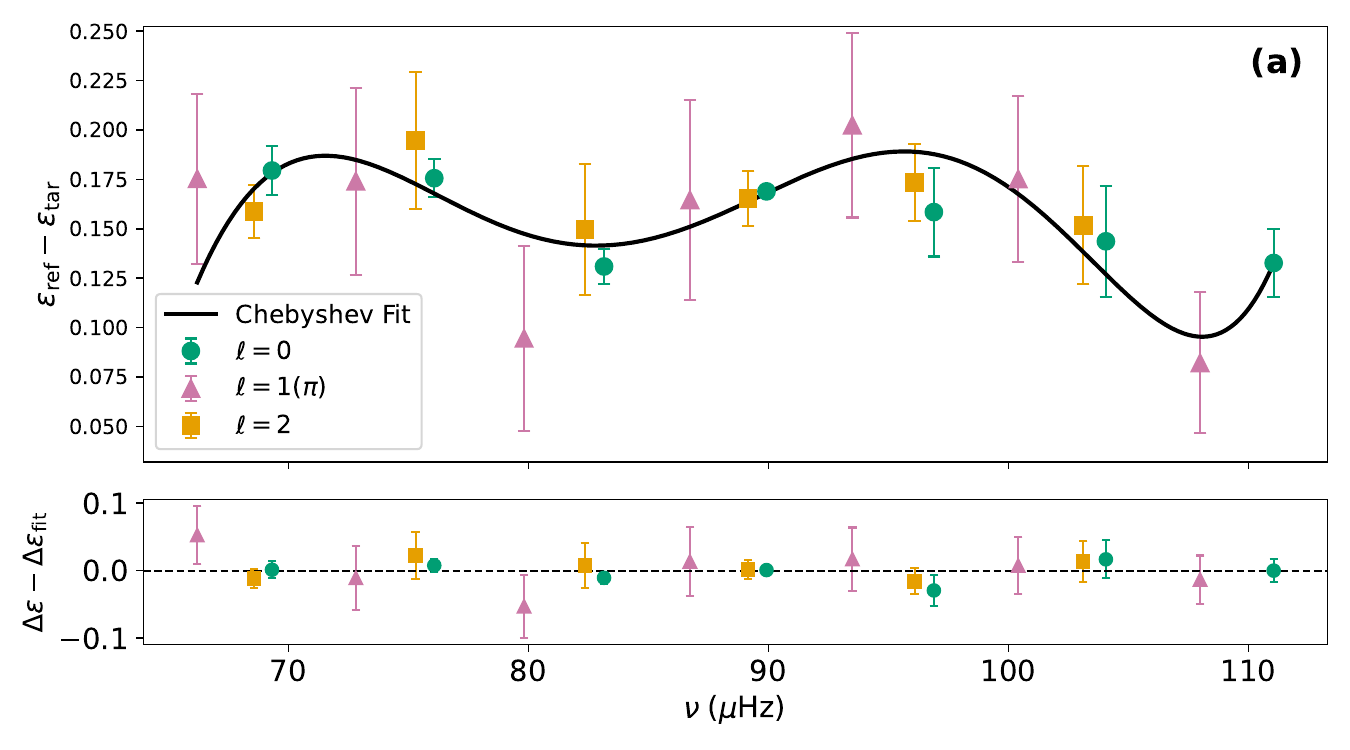}
        \phantomcaption
        \label{fig:edkckep1}
    \end{subfigure}
    \hfill
    \begin{subfigure}{0.48\linewidth}
        \centering
        \includegraphics[width=\linewidth]{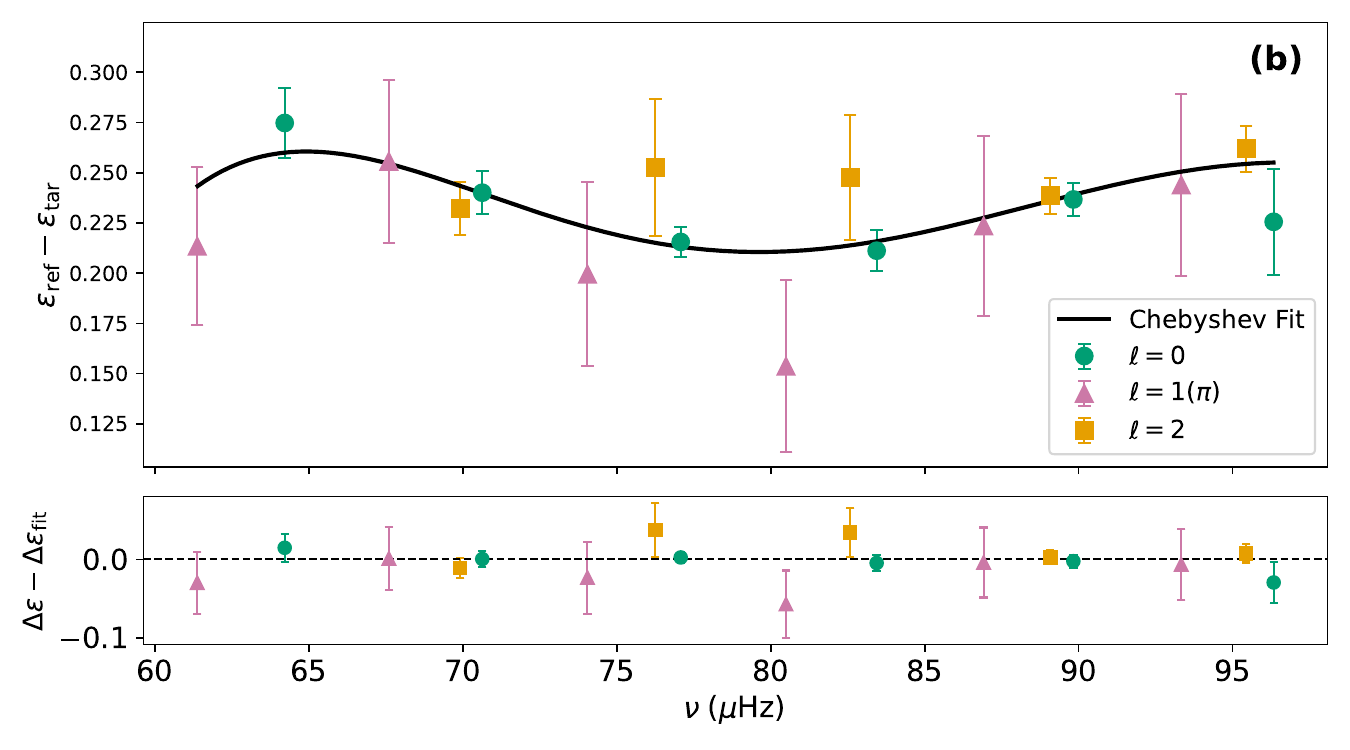}
        \phantomcaption
        \label{fig:edkckep2}
    \end{subfigure}
    \caption{$\epsilon_p$-difference diagrams between: \textbf{(a)} $\kappa$~Cyg and HD~226808 and \textbf{(b)} $\kappa$~Cyg and HD~181827.}
    \label{fig:edkckep}
\end{figure*}

To diagnose potential differences in the interior structures of these stars, we applied the $\epsilon_p$-difference technique on a star-versus-star basis, between $\kappa$~Cyg and the two \kepler{} stars. To our knowledge, this is the first time, the $\epsilon_p$-difference technique has been applied in this fashion, differentially between two real stars. We treat $\kappa$~Cyg as our reference star and each of the \kepler{} stars as targets. The diagrams can be seen in \figref{fig:edkckep}. 

In contrast to the star-versus-model comparison, both the reference and target frequency sets carry observational uncertainties. This naturally leads to smaller $\chi^2$ values and therefore somewhat reduced sensitivity to subtle $\ell$-dependent signatures. Nevertheless, we apply the same statistical framework to test for $\ell$-dependence. For HD~226808, the nested-model comparison yields $\Delta\chi^2$ = 0.27 ($p = 0.876$), F-test $p = 0.856$, $\Delta$AIC = -3.73 and $\Delta$BIC = -5.73, all consistent with an $\ell$-independent $\epsilon_p$ difference. Similarly, for HD~181827, we obtain $\Delta\chi^2$ = 2.01 ($p = 0.366$), F-test $p = 0.301$, $\Delta$AIC = -1.99 and $\Delta$BIC = -3.66, consistent with $\ell$-independent $\epsilon_p$ difference. Hence, we conclude that within the observational uncertainties, we do not detect a clear $\ell$-dependence in the inter-star $\epsilon_p$ differences for either target, indicating that their interior structures are broadly comparable to that of $\kappa$~Cyg. \ac{Since the models were shown in \secref{kappa-modelling} to misrepresent the interior structure of $\kappa$~Cyg, it follows that the same models should fail comparably for HD~181827.} Therefore, it is not a surprise that the modelled radius of HD~181827 is higher than the interferometric one.

\section{Conclusions}
\label{conclusions}
We have performed the first detailed characterization of the secondary clump star, $\kappa$~Cyg, by combining high-precision \tess{} asteroseismology with visible-wavelength interferometry using the PAVO beam combiner at the CHARA Array. This dual approach allowed for a stringent test of stellar evolutionary models in the core helium-burning phase, which remains a regime of significant theoretical uncertainty. Our main findings are:
\begin{itemize}
    \item Squared-visibility measurements beyond the first null allow a direct fit to the four-parameter non-linear limb-darkening law. We found that the 3D hydrodynamic STAGGER model provides a limb-darkened radius for $\kappa$~Cyg by approximately 2.1~per cent higher than that derived from a direct fit. Using synthetic spectra from PHOENIX atmospheric models and spectrophotometric magnitudes of $\kappa$~Cyg, we calculated $F_{\rm bol}$ by fitting the spectral energy distribution function. We noticed that using Pickles atlas of stellar spectra gives $F_{\rm bol}$ lower by about 2.7~per cent.
    \item Using 16 sectors of TESS photometry, we identified clear solar-like oscillations in the power spectrum of $\kappa$~Cyg. By employing stretched-period and stretched-frequency \echelle{} diagrams, we decoupled the mixed dipole modes to extract $\pi$-modes, which are sensitive to the acoustic cavity and provide a robust seismic probe that is independent of the complex core-coupling physics.
    \item We compared the observed oscillation frequencies of $\kappa$ Cyg against two extensive MESA grids using either predictive mixing or exponential overshooting. Both grids systematically overestimated the stellar radius compared to our interferometric measurement. Although the overshooting grid performed marginally better, it yielded a median radius approximately 3.2~per cent higher than the observed value. 
    \item By repeating our asteroseismic analysis for two \kepler{} secondary clump stars, HD~226808 and HD~181827, we investigated the radius discrepancy observed in $\kappa$~Cyg. Although the models do reproduce the \gaia{} parameters of HD~226808 accurately, the interferometric radius of HD~181827 is consistently overestimated, as in the case of $\kappa$~Cyg.
    \item There is a clear internal inconsistency in current 1D models of secondary clump stars: models that reproduce the envelope-dominated p-mode frequencies systematically predict stellar radii larger than the interferometric measurement. The largely geometric nature of the interferometric radius implies that the models, rather than the observations, are incorrect. 
    \ac{The uncertainty on the interferometric radius is dominated by contributions from $\theta_{\rm LD}$ and the parallax $\varpi$ (\eqnref{eq:radius-parallax}). The $\theta_{\rm LD}$ error budget includes both statistical uncertainties and systematics from calibration and the adopted limb-darkening (LD) prescription. However, because our observations extend well beyond the first null of the visibility curve, calibration systematics are not expected to dominate since the location of the first null is precisely known \citep{Tayar2022}. Instead, the primary systematic arises from the LD treatment: comparing our direct fit and STAGGER-based fit yields a maximum LD difference of $\sim2.1$ per cent (~\tabref{tab:interResults}), but the latter yields a poor fit to the second lobe of stellar visibilities. Inflating ($\sigma_\varpi$) using the RUWE-based prescription of \citet{ElBadry2025} increases the radius uncertainty from 0.10 to 0.16 R$_\odot$; even under this conservative treatment, the model discrepancy of ($\sim0.28$ R$_\odot$) exceeds the total uncertainty budget and cannot be fully explained by systematics in either $\theta_{\rm LD}$ or $\varpi$.
}

\end{itemize}

The last conclusion is further supported by the fact that the same models also underpredict the dipole-mode period spacing $\Delta\Pi_1$. The failure to reproduce both the global radius and the buoyancy-related asteroseismic diagnostics indicates that the internal structure of the models---most likely the treatment of core boundary mixing during and prior to core-helium burning---is not fully realistic. Although we isolated the pure p ($\pi$) modes to minimise sensitivity to the core structure and core mixing, the results indicate that some residual influence from the core remains. While the models employed here are representative of the current state of the art in 1D stellar evolution, aided by interferometry, $\kappa$~Cyg acts as a benchmark to demonstrate that matching envelope-sensitive asteroseismic constraints alone is insufficient to guarantee a correct global or core structure. Future work will focus on expanding the list of secondary clump stars with interferometric data to reconcile the tension between asteroseismic predictions and direct interferometric radii.

\section*{Acknowledgements}
AC, TRB, LSS and CLC acknowledge support from the Australian Research Council through Laureate Fellowship FL220100117. DH acknowledges support from the National Aeronautics and Space Administration (80NSSC22K0781) and the Australian Research Council (FT200100871).

This work uses observations obtained with the Georgia State University Center for High Angular Resolution Astronomy Array at Mount Wilson Observatory. The array is supported by the National Science Foundation under Grant No. AST-1211929 and AST-1411654. Institutional support has been provided from the GSU College of Arts and Sciences and the GSU Office of the Vice President for Research and Economic Development. We have used data collected by the \tess{} and \kepler{} missions and obtained from the MAST data archive at the Space Telescope Science Institute (STScI). Funding for the \kepler{} mission is provided by the NASA Science Mission Directorate, and for the \tess{} mission by the NASA Explorer Programme. STScI is operated by the Association of Universities for Research in Astronomy, Inc., under NASA contract NAS 5-26555. We use data from the \gaia{} mission, funded by the European Space Agency. This paper has made use of the SIMBAD\footnote{\url{https://simbad.u-strasbg.fr/simbad/sim-basicIdent=m33&submit=SIMBAD+search}} database \citep{Wegner2000}, operated at CDS, Strasbourg, France.

\textit{Software:}  We have extensively used several public \textsc{Python} packages: \textsc{NumPy} \citep{Harris2020}; \textsc{Matplotlib} \citep{Hunter2007}; \textsc{SciPy} \citep{Virtanen2020}; \textsc{Astropy} \citep{astropy:2013, astropy:2018, astropy:2022}; \textsc{Pandas} \citep{mckinney2010, reback2020pandas}; \textsc{Emcee} \citep{Foreman2013}; \textsc{lightkurve} \citep{LK2018}.

This research has used the open-source stellar evolutionary code, Modules for Experiments in Stellar Astrophysics  (\textsc{Mesa}, version r24.03.1:\citealt{Paxton2011, Paxton2013, Paxton2015, Paxton2018, Paxton2019, Jermyn2023}). The equation of state (EOS) implemented \textsc{Mesa} is constructed from a composite of several sources, including OPAL \citep{Rogers2002}, SCVH \citep{Saumon1995}, FreeEOS \citep{Irwin2012}, HELM \citep{Timmes2000}, PC \citep{Potekhin2010}, and Skye \citep{Jeremyn2021}. Radiative opacities are taken primarily from the OPAL tables \citep{Iglesias1993, Itoh1996}, supplemented at low temperatures by \citet{Ferguson2005} and extended to the high-temperature, Compton–scattering–dominated regime using the prescription of \citet{Poutanen2017}. Electron conduction opacities follow \citet{Cassisi2007} and \citet{Blouin2020}. Nuclear reaction rates are adopted from the JINA REACLIB database \citep{Cyburt2010}, NACRE \citep{Angulo1999}, along with tabulated weak interaction rates from \citet{Fuller1985, Oda1994, Langanke2000}. Coulomb screening is treated using the formulation of \citet{Chugunov2007}, while thermal neutrino energy-loss rates are taken from \citet{Itoh1996}.

\section*{Data Availability}
The \tess{} and \kepler{} data underlying this article are available at the MAST Portal (Barbara A. Mikulski Archive for Space Telescopes), at \hyperlink{https://mast.stsci.edu/portal/Mashup/Clients/Mast/Portal.html}{https://mast.stsci.edu/portal/Mashup/Clients/Mast/Portal.html}. The stellar visibilities and stellar models (\textsc{MESA} and \textsc{GYRE} inlists) are available under this DOI: \href{https://doi.org/10.5281/zenodo.18971442}{https://doi.org/10.5281/zenodo.18971442}.


\bibliographystyle{mnras}
\bibliography{main}



\appendix
\section{LD-4 coefficients}
\begin{figure*}
    \centering
    \includegraphics[width=\linewidth]{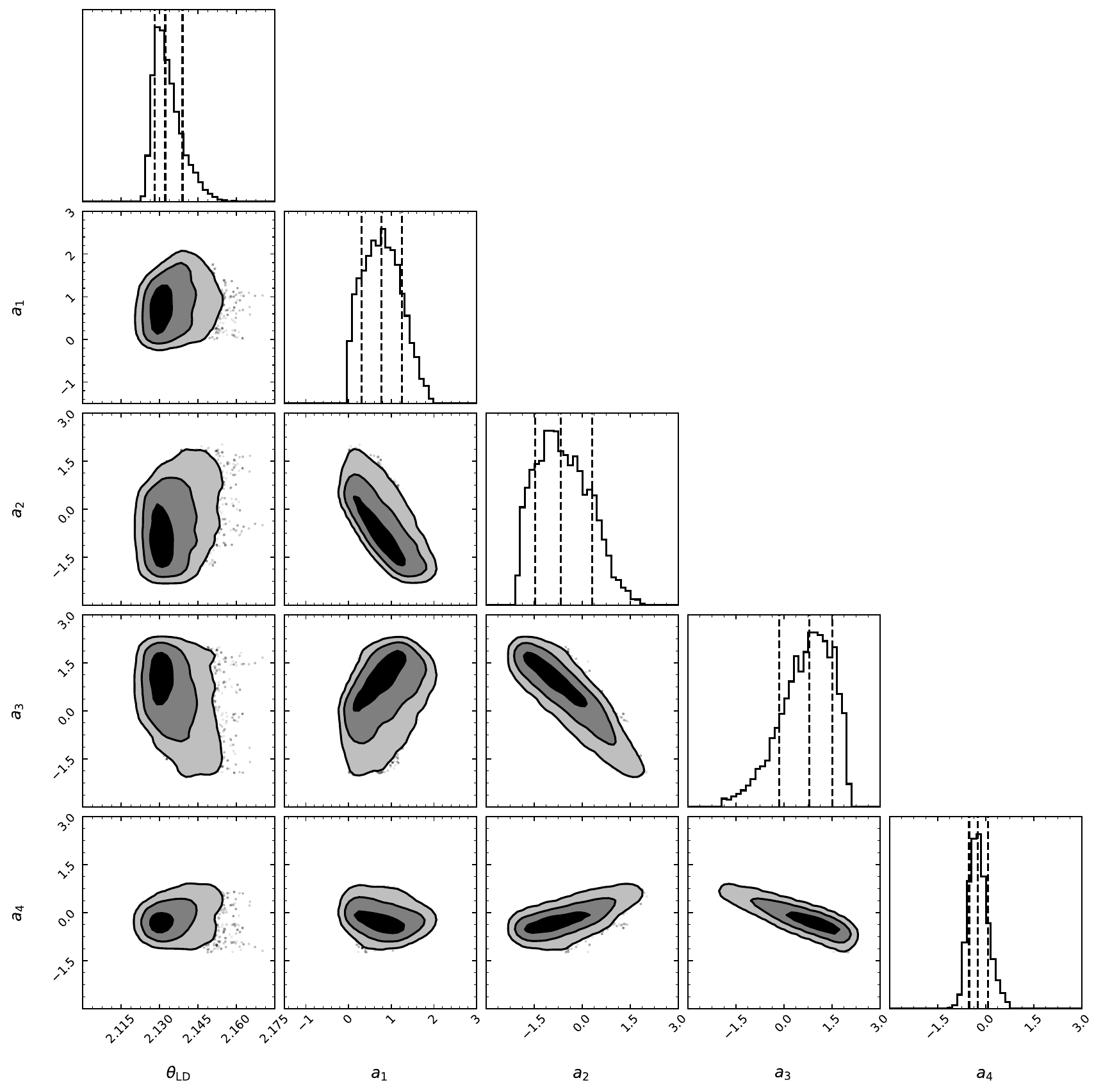}
    \caption{Corner plot of $\theta_{\rm LD}$ and LD coefficients with contour lines representing the 1, 2 and 3 $\sigma$ credible regions.}
    \label{fig:cornera}
\end{figure*}

\begin{table*}
    \centering
    \begin{tabular}{lcccccccccccc}
    \hline
    $\mu$ & 0.01 & 0.05 & 0.12 & 0.22 & 0.34 & 0.47 & 0.60 & 0.72 & 0.83 & 0.92 & 0.97  \\
    \hline
    $\dfrac{I(\mu)}{I(1)}$ & $0.45_{-0.12}^{+0.08}$ & $0.53_{-0.08}^{+0.05}$ & $0.60_{-0.05}^{+0.03}$ & $0.66_{-0.03}^{+0.02}$ & $0.73_{-0.01}^{+0.01}$ & $0.79_{-0.01}^{+0.00}$ & $0.84_{-0.00}^{+0.00}$ & $0.89_{-0.00}^{+0.00}$ & $0.94_{-0.00}^{+0.00}$ & $0.97_{-0.00}^{+0.00}$ & $0.99_{-0.00}^{+0.00}$ \\
    \hline
    \end{tabular}
    \caption{Reconstructed centre-to-limb variation in intensity from direct LD fit to \eqnref{eq:visibility_polynomial} for $\kappa$ Cyg.}
    \label{tab:clv}
\end{table*}

As expected, our fit does not constrain the 4 LD coefficients ($a_k$) of \eqnref{eq:visibility_polynomial} (see \figref{fig:cornera}). So, we report the reconstructed centre-to-limb intensity values of $\kappa$ Cyg from its squared visibilities in \tabref{tab:clv}.

\section{Corner Plot}
\begin{figure*}
    \centering
    \includegraphics[width=\linewidth]{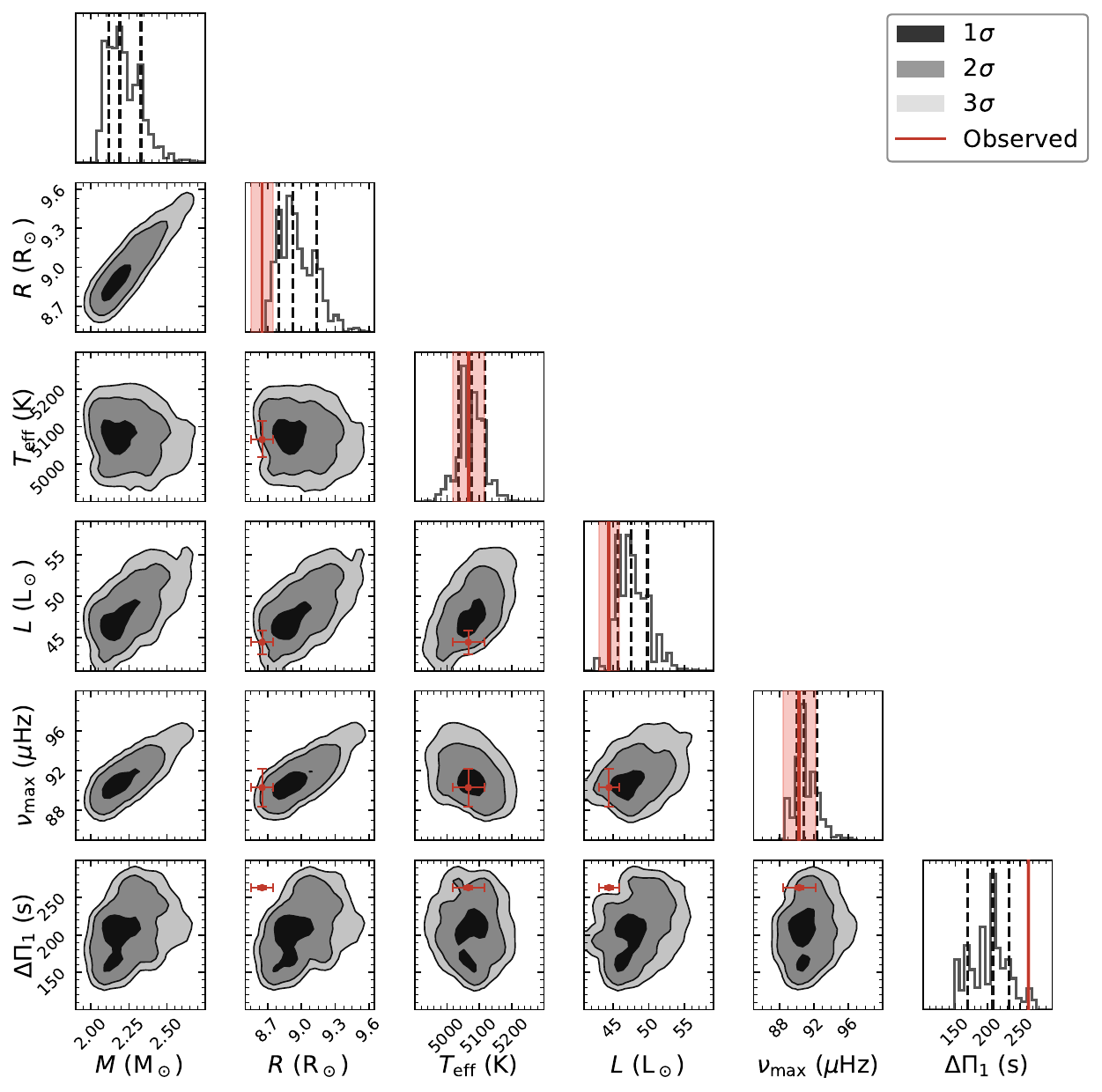}
    \caption{Corner plot for $\kappa$~Cyg using grid-OS (overshooting). The grayscale panels show the normalised probability distribution calculated using a kernel density estimation (KDE). The contour lines represent the 1, 2 and 3 $\sigma$ credible regions. The diagonal panels show the normalised probability distributions with the median and 1 $\sigma$ quantiles shown in vertical dashed lines. The observed values are shown in red patches.}
    \label{fig:corner}
\end{figure*}
Since the grid-OS using exponential overshooting performs better than grid-PM for $\kappa$~Cyg, we present its corresponding corner plot in \figref{fig:corner}.


\bsp	
\label{lastpage}
\end{document}